\documentclass[twocolumn,floatfix,aps,prb,superscriptaddress,10pt]{revtex4-1}

\usepackage[urlcolor=blue, colorlinks=true, citecolor=blue,linkcolor=blue]{hyperref}
\usepackage[utf8]{inputenc}
\usepackage[normalem]{ulem}
\usepackage{float, graphicx, array, multirow, tabularx}
\usepackage{amsmath, amssymb}
\usepackage{color, xcolor}
\newcommand{\ZZ}{\mathbb{Z}}
\newcommand{\dis}{\overline}
\newcommand{\Ins}{{\rm I}}
\newcommand{\vk}{\mathbf{k}}
\renewcommand{\vr}{\mathbf{r}}
\newcommand{\ket}[1]{|#1\rangle}

\newcommand{\bl}[1]{ {\color{blue}#1} }

\begin{document}

\title{Disordered topological crystalline phases}

\author{Adam Yanis Chaou}
\affiliation{Dahlem Center for Complex Quantum Systems and Institut f\"ur Physik, Freie Universit\"at Berlin, Arnimallee 14, D-14195 Berlin, Germany}

\author{Mateo Moreno-Gonzalez}
\affiliation{Institute for theoretical physics, Universit\"at zu K\"oln, Z\"ulpicher Str. 77a, 50937 Cologne, Germany}

\author{Alexander Altland}
\affiliation{Institute for theoretical physics, Universit\"at zu K\"oln, Z\"ulpicher Str. 77a, 50937 Cologne, Germany}

\author{Piet W.\ Brouwer}
\affiliation{Dahlem Center for Complex Quantum Systems and Institut f\"ur Physik, Freie Universit\"at Berlin, Arnimallee 14, D-14195 Berlin, Germany}

\begin{abstract}
The imposition of crystalline symmetries is known to lead to a rich variety of insulating and superconducting topological phases. These include higher-order topological phases and obstructed atomic limits with and without filling anomalies. We here comprehensively classify such topological crystalline phases (TCPs) with mirror, twofold rotation, and inversion symmetries in the presence of disorder that preserves the crystalline symmetry on average. We find that the inclusion of disorder leads to a simplification of the classification in comparison to the clean case. We also find that, while clean TCPs evade a general bulk-boundary principle, disordered TCPs admit a complete bulk-boundary correspondence, according to which (bulk) topological phases are topologically equivalent if and only if they have the same anomalous boundary states and filling anomaly. We corroborate the stability of disordered TCPs by way of field-theoretic, numerical and symmetry-based analyses in various case studies. While the boundary signatures of most disordered TCPs are similar to their clean counterparts, the addition of disorder to certain mirror-symmetric TCPs results in novel higher-order statistical topological phases, in which zero-energy hinge states have critical wavefunction statistics, while remaining protected from Anderson localization.
\end{abstract}

\maketitle

\section{Introduction} 
\label{sec:intro}

Topological band insulators and superconductors satisfy a bulk-boundary correspondence: A nontrivial bulk topology implies the existence of ``anomalous'' states at the system boundary and vice versa.\cite{hasan2010,qi2011,bernevig2013,ando2015} Examples of such anomalous boundary states include the chiral edge modes of the quantized Hall effect or the helical edge modes of the quantum spin-Hall effect. A complete classification of topological insulators and superconductors (hereafter simply referred to as ``topological phases'') and their associated anomalous boundary states --- protected by the local ``tenfold-way'' symmetries of time-reversal and particle-hole symmetry and their product\cite{altland1997} --- is given by the ``periodic table of topological insulators and superconductors''.\cite{schnyder2008, schnyder2009, kitaev2009}

If crystalline symmetries,\cite{fu2007b,fu2011, ando2015, turner2012, chiu2013, morimoto2013, shiozaki2014, trifunovic2017} such as mirror, rotation, or inversion symmetry, are also imposed, the topology of the insulating bulk may also manifest through higher-order boundary states.\cite{benalcazar2017, benalcazar2017b, peng2017, song2017, schindler2018, langbehn2017, geier2018, fang2019, khalaf2018, khalaf2018b, trifunovic2019, trifunovic2021} These states are similar to the anomalous boundary states of tenfold-way topological insulators, except that they exist on crystal hinges or corners, rather than on the entire surface. Unlike their non-crystalline counterparts, topological crystalline phases (TCPs) do not have a complete bulk-boundary correspondence: There exist TCPs with nontrivial bulk topology, but no boundary signatures.\cite{bradlyn2017} Classifications of TCPs that account for the possible presence of higher-order boundary states exist,\cite{geier2018,khalaf2018b,trifunovic2019} but they are more involved than those of the tenfold-way classes.

In this article we address TCPs in the presence of crystalline-symmetry-breaking disorder, where the disorder preserves the symmetry on average. We consider three crystalline symmetries: mirror, twofold rotation, and inversion symmetries. For these symmetries, we find that not all topological distinctions between crystalline phases persist in the presence of disorder. This ``blurring'' of topological distinctions leads to a {\em simplification} of the classification of disordered TCPs in comparison to clean TCPs, which is such that disordered TCPs again have a {\em complete bulk-boundary correspondence}: For disordered TCPs, a nontrivial bulk topology is uniquely tied to anomalous boundary states or to a filling anomaly, a deviation from strict charge neutrality for a complete filling of the valence band in a symmetry-compatible crystal.\cite{benalcazar2019} We also find that in certain symmetry classes, disordered TCPs may have second-order hinge modes that are {\em critical}, with power-law correlations at large distances. Such phenomenology is unique to disordered systems and is not found in clean TCPs.

There exists a wealth of evidence in the literature that crystalline topology can survive the breaking of translation symmetry: Numerous case studies in amorphous,\cite{corbae_amorphous_2023, wang_structural-disorder-induced_2021, agarwala_higher-order_2020, peng_density-driven_2022} disordered crystalline,\cite{loio_third-order_2024,li_topological_2020,yang_higher-ordpeer_2021,su_disorder_2019,wang_disorder-induced_2020,hu_disorder_2021,coutant_robustness_2020,zhang_experimental_2021,franca_phase-tunable_2019,shen_disorder-induced_2024,song_delocalization_2021, wang_anderson_2024} or quasi-crystalline systems\cite{peng_higher-order_2021,varjas_topological_2019,velury_topological_2021} have demonstrated that anomalous higher-order signatures do not always require a pristine setting. Furthermore, amorphous systems with average crystalline symmetries can host statistical topological phases\cite{spring_amorphous_2021, spring_isotropic_2023} and amorphous obstructed insulators,\cite{marsal_obstructed_2023} where boundary manifestations of the topology affect the entire boundary, rather than hinges or corners only. Our work offers a comprehensive understanding of when TCPs survive symmetry-breaking disorder and of the fate of their boundary signatures.

Of the three crystalline symmetries we consider in this article, inversion always acts non-locally on the crystal surface. In the absence of disorder, this a prerequisite for the existence TCPs without boundary states, phases known as ``obstructed atomic-limits''.\cite{bradlyn2017} Although they do not have anomalous boundary states, atomic-limit phases may have a filling anomaly,\cite{benalcazar2019} which manifests by way of fractional charges at crystal corners.\cite{benalcazar2017,benalcazar2019,vanmiert2017,vanmiert2018,rhim2017,hughes2011,watanabe2020,ren2021,trifunovic2020,schindler2020,kooi2021} It is because of the existence of obstructed atomic limit phases {\em without} a filling anomaly --- and, hence, without any anomalous boundary signature --- that TCPs only have a partial bulk-boundary correspondence. Considering the effect of disorder on obstructed atomic-limit phases, we find that the obstructed atomic-limit phases that are trivialized by disorder are precisely those that do not have a filling anomaly. Our conclusion that disordered TCPs have a complete bulk-boundary correspondence is a consequence of this observation.

Unlike inversion symmetry, for mirror symmetry (in $2d$ and $3d$) and twofold rotation symmetry (in $3d$) there always exist points at the crystal surface at which the symmetry acts locally. The existence of such invariant points on the crystal surface ensures the presence of boundary states and, hence, rules out obstructed atomic-limit phases protected by mirror or rotation symmetry.\cite{trifunovic2019} As disorder breaks the protecting mirror or rotation symmetry, it may trivialize certain higher-order boundary states along high-symmetry hinges or corners. The simplification of the classification of disordered TCPs in comparison to clean TCPs that we report here originates from the trivialization of such higher-order TCPs, as well as the trivialization of atomic-limits without a filling anomaly.

An interesting scenario arises in mirror-symmetric higher-order TCPs in $3d$ with counter-propagating modes at a mirror-symmetric hinge. In general, such modes undergo Anderson localization if disorder is added. However, in certain cases the fact that the disorder respects mirror symmetry on average can allow the hinge modes to avoid Anderson localization: Instead of being exponentially localized, they become critical, acquiring power-law correlations at large distances. Similar states also appear at the boundary of two-dimensional ``statistical topological insulators''.\cite{fulga2014} We hence refer to such disordered topological phases as a ``second-order statistical topological insulator''.

The remainder of this article is organized as follows: In Sec.\ \ref{sec:casestudy}, we use three examples to review the topology of higher-order boundary states and obstructed atomic limits and to illustrate how disorder impacts the classification of TCPs. The examples considered in Sec.\ \ref{sec:casestudy} illustrate the three main conclusions of this article: The simplification of the classification in the presence of disorder, the restoration of the bulk-boundary correspondence, and the appearance of critical hinge states as a possible boundary signature. Details of the proof of the bulk-boundary correspondence for higher-order TCPs with mirror or twofold rotation symmetry are given in App.\ \ref{app:BB}. The methodology used to classify disordered obstructed atomic limits and show the triviality of disordered atomic limits without a filling anomaly is laid out in App.\ \ref{app:dis_OAL}. Details concerning the second-order statistical topological insulator phase are given in App.\ \ref{app:statistical}. Classification results for all tenfold-way classes, with an additional (statistical) mirror, twofold rotation, or inversion symmetry then follow in Sec.\ \ref{sec:classification} and App.\ \ref{app:classification}. Section \ref{sec:microscopics} contains field-theoretic and numerical analyses of two concrete microscopic examples, one demonstrating the robustness of chiral hinge modes to disorder and, the other, corroborating the bulk-boundary correspondence of disordered TCPs. We conclude with a brief discussion in Sec.\ \ref{sec:conc}. 

\section{Case studies}
\label{sec:casestudy}

We begin with an illustrative discussion of how disorder affects the topological classification of three examples of topological crystalline phases. The first two examples feature mirror-symmetric insulators and superconductors, the third considers an inversion-symmetric insulator. The first example features zero-energy corner modes at a mirror-symmetric corner, which are gapped out by disorder, whereas the second example considers a statistical second-order topological insulator with critical hinge modes. The third example illustrates how disorder trivializes precisely those obstructed atomic-limit TCPs that do not have a filling anomaly.

For these three examples, we will encounter {\em bulk topological invariants} and {\em boundary invariants}. The boundary invariants characterize configurations of anomalous boundary states in a manner that is robust to arbitrary perturbations at the crystal boundary, but not in the bulk. The bulk invariant characterizes topological equivalence classes of TCPs: If two TCPs have the same bulk invariant, they can be continuously deformed into each other while remaining insulating in the bulk throughout the deformation. Since disorder breaks the crystalline symmetry, we find that we cannot define the bulk invariant of a disordered TCP using the bulk band structure. Instead, we make use of the fact that disordered TCPs have a bulk-boundary correspondence, so that knowledge of the anomalous boundary states is sufficient to decide on the possibility to continuously deform two bulk TCPs into each other and vice versa.

\subsection{Mirror-symmetric TCP with corner states}
\label{sec:Ex1}

The first example features a two-dimensional TCP with zero-energy corner states. The TCP is in class AIII and has an additional mirror symmetry. Gapped band structures $H(k_x,k_y)$ in class AIII satisfy the symmetry constraint
\begin{equation}
  H(k_x,k_y) = - \sigma_3 H(k_x,k_y) \sigma_3,
  \label{eq:chiral}
\end{equation}
where  $\sigma_3$ is a Pauli matrix that represents chiral conjugation. We additionally impose a mirror symmetry $y \to -y$ that commutes with $\sigma_3$,
\begin{equation}
  H(k_x,k_y) = \tau_3 H(k_x,-k_y) \tau_3,
  \label{eq:mirror}
\end{equation}
where $\tau_3$ is a Pauli matrix acting on different degrees of freedom than $\sigma_3$.  The eigenvalues $\sigma$ and $\tau$ of $\sigma_3$ and $\tau_3$ are referred to as ``chirality'' and ``mirror parity'', respectively. We refer to the symmetry class of Hamiltonians satisfying Eqs.\ (\ref{eq:chiral}) and (\ref{eq:mirror}) as class $\mbox{AIII}^{{\cal M}_+}$, where the superscript ${\cal M}_+$ indicates the presence of the additional mirror symmetry that commutes with the chiral conjugation $\sigma_3$. Below, we first discuss the topological classification of non-disordered two-dimensional TCPs in class $\mbox{AIII}^{{\cal M}_+}$ and then turn to the effect of disorder.

{\em Classification of bulk band structure.---} At the high-symmetry lines $k_y = 0$/$\pi$, the mirror symmetry (\ref{eq:mirror}) imposes that $H(k_x,k_y)$ is the diagonal sum of blocks $H_{\tau}(k_x,0/\pi)$ with mirror parity $\tau = \pm$. The topological invariants of the bulk band structure may be obtained from the winding numbers $W_{\tau}(0/\pi)$ of these diagonal blocks, seen as functions of $k_x$. The four winding numbers satisfy the constraint $W_{+}(0) + W_{-}(0) = W_{+}(\pi) + W_{-}(\pi)$. Band structures with $W_{\tau}(0) = W_{\tau}(\pi)$, $\tau = \pm$, are weak topological phases, which are topologically equivalent to stacks of one-dimensional band structures with an additional mirror symmetry. Here we will restrict ourselves to band structures without weak topology and therefore impose that $W_{+}(0) = W_{-}(0) = 0$. Such band structures are described by the integer topological invariant,
\begin{equation}
  Q_{\rm bulk} = W_-(\pi). \label{eq:QbulkAIII}
\end{equation}

\begin{figure}
  \centering
  \includegraphics[width=0.65\columnwidth]{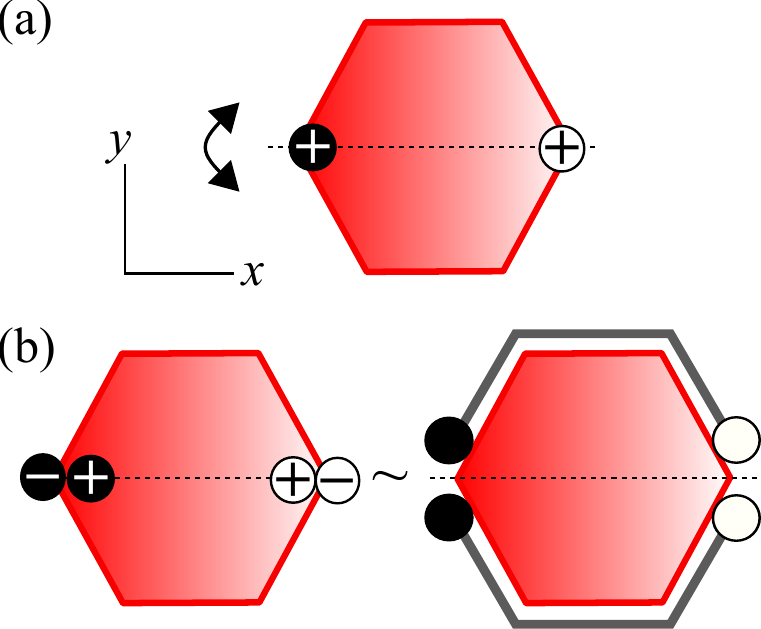}
  \caption{ (a) A second-order topological insulator in class $\mbox{AIII}^{{\cal M}_+}$ with a mirror symmetry $y \to -y$ and with zero-energy states at mirror-symmetric corners. Such zero-energy states have well-defined chirality (black vs.\ white circles) and mirror parity $\pm$. (Mirror parity is not well-defined for zero-energy states away from mirror-symmetric corners.) (b) An insulator with pairs of corner states with the same chirality, but opposite mirror parity, can be continuously deformed to an insulator without zero-energy corner states without closing the bulk gap. Such a deformation involves a ``boundary decoration'', a perturbation that extends along the entire crystal boundary.}    \label{fig:layer}
\end{figure}

{\em Boundary classification.---} The anomalous boundary states associated with such insulating band structures are zero-energy states localized at mirror symmetric corners.\cite{langbehn2017,geier2018,trifunovic2019} Such corner states have well-defined chirality and mirror parity.
Denoting the number of zero-energy corner states with chirality $\sigma$ and mirror parity $\tau$ by $n_{\sigma\tau}$, the differences
\begin{equation}
  N_{{\rm corner},\tau} = n_{+\tau} - n_{-\tau} \label{eq:NcornerAIII}
\end{equation}
are robust to local perturbations at the corner that respect the chiral and mirror symmetries. Configurations of corner states with nonzero $N_{{\rm corner},\tau}$ are called ``anomalous'', because they cannot be obtained as the eigenstates of a stand-alone lattice Hamiltonian localized at the corner and obeying chiral and mirror symmetries. 

The integers $N_{{\rm corner},+}$ and $N_{{\rm corner},-}$ are not individually robust to ``boundary decorations'', perturbations that extend along the entire crystal boundary, but do not affect the bulk of the crystal. Figure \ref{fig:layer} shows an example, how a boundary decoration can be used to deform a crystal with $N_{{\rm corner},+} = N_{{\rm corner},-} = 1$ to a crystal with $N_{{\rm corner},+} = N_{{\rm corner},-} = 0$ without closing the bulk gap. In general, boundary decorations may simultaneously change $N_{{\rm corner},+}$ and $N_{{\rm corner},-}$ by the same (arbitrary) amount, so that only the difference
\begin{equation}
  Q_{\rm boundary} = N_{{\rm corner},+} - N_{{\rm corner},-} 
\end{equation}
is a true topological invariant that is robust to all perturbations that do not close the bulk gap. The boundary invariants $N_{{\rm corner},+}$ and $N_{{\rm corner},-}$ are therefore referred to as ``extrinsic'', whereas $Q_{\rm boundary}$ is called ``intrinsic''.\cite{geier2018}

{\em Bulk-boundary correspondence.---} The intrinsic boundary invariant $Q_{\rm boundary}$ is in one-to-one correspondence with the bulk invariant $Q_{\rm bulk}$,\cite{trifunovic2019}
\begin{equation}
  Q_{\rm bulk} = Q_{\rm boundary}.
  \label{eq:Qbulk}
\end{equation}
This relation is known as the ``bulk-boundary correspondence''. 

{\em Disorder: Boundary classification.---} We consider disorder that respects the chiral symmetry (\ref{eq:chiral}) and that obeys the mirror symmetry (\ref{eq:mirror}) on average. Since disorder always locally breaks mirror symmetry, it blurs the distinction between corner states with even and with odd mirror parity, so that of the two extrinsic boundary invariants $N_{{\rm corner},+}$ and $N_{{\rm corner},-}$, only the sum
\begin{equation}
  \dis N_{\rm corner} = 
  N_{{\rm corner},+} + N_{{\rm corner},-}
\end{equation}
remains well-defined in the presence of disorder. However, because the disorder respects the mirror symmetry on average, decorations on mirror-related boundary segments must have the same topology, so that boundary decorations can only change the total number of corner states $\dis N_{\rm corner}$ by an even amount. This means that in the presence of disorder only the parity
\begin{align}
  \dis Q_{\rm boundary} =&\, 
  (N_{{\rm corner},+} + N_{{\rm corner},-}) \mod 2 \nonumber \\ =&\,
  Q_{\rm boundary} \mod 2
  \label{eq:Qcorner}
\end{align}
remains as an intrinsic boundary invariant.

\begin{figure}
  \centering
  \includegraphics[width=0.92\columnwidth]{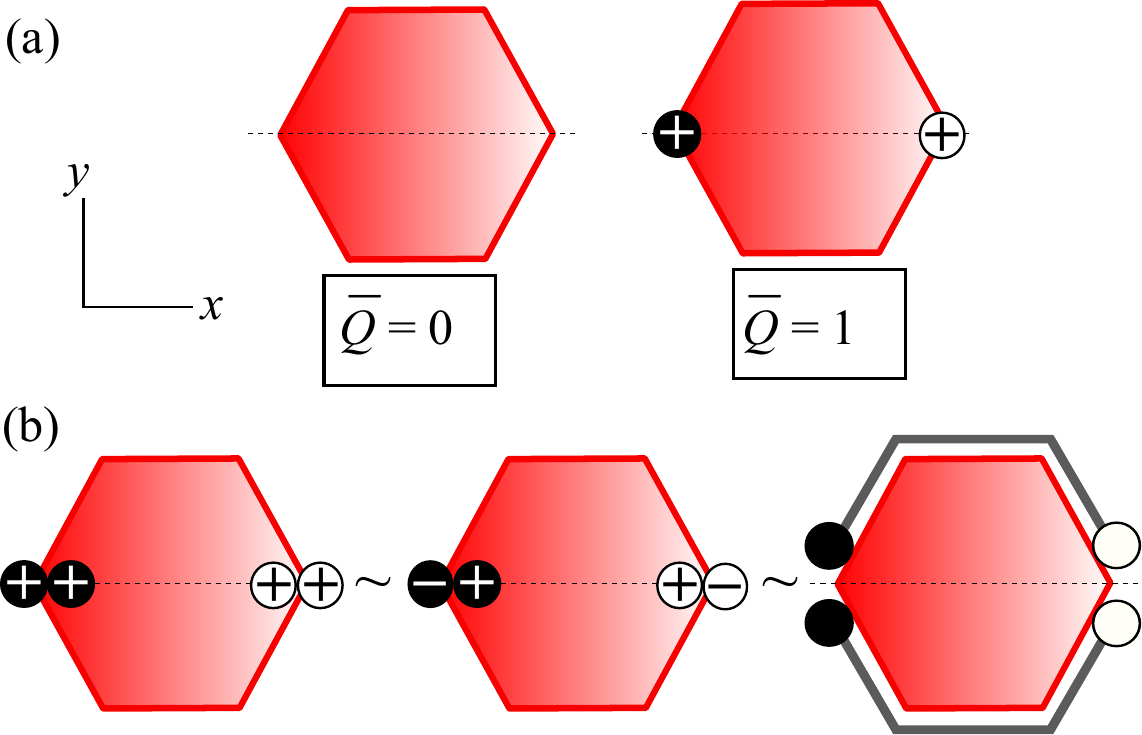}
  \caption{(a) Schematic representations of the topological equivalence classes of disordered TCPs in class $\mbox{AIII}^{{\cal M}_{+}}$, corresponding to bulk topological invariants $\dis{Q}_{\rm bulk} = 0$ and $\dis{Q}_{\rm bulk} = 1$. (b) Illustration of the $\ZZ_2$ group structure of the topology for disordered TCPs in class $\mbox{AIII}^{{\cal M}_{+}}$: The direct sum of two TCPs with $\dis{Q}_{\rm bulk} = 1$ is topologically trivial, {\em i.e.}, it can be continuously deformed to the trivial phase without closing the bulk mobility gap. The first step of the deformation makes use of the fact that in the presence of disorder it is no longer possible to distinguish between corner states of even and odd mirror parity. The second step involves a boundary deformation and is the same as in Fig.\ \ref{fig:layer}(b).}
    \label{fig:generator}
\end{figure}

\emph{Disorder: Bulk-boundary correspondence.---}Disordered TCPs with different $\dis{Q}_{\rm boundary}$ are necessarily topologically distinct from a bulk perspective. For a bulk-boundary correspondence to exist, the opposite must also be true: Disordered TCPs with the same intrinsic boundary invariant can be continuously deformed into each other without closing the bulk mobility gap. In App.\ \ref{app:BB} we verify that this is indeed the case for TCPs in class $\mbox{AIII}^{{\cal M}_{+}}$. The arguments make use of the ``layer representation'' of a higher-order TCP\cite{isobe2015,fulga2016,huang2017,khalaf2018,trifunovic2019} and can easily be generalized to other higher-order topological phases. Having established that there is a bulk-boundary correspondence for second-order topological phases and since there are no anomalous first-order boundary states in this symmetry class, we may therefore define the bulk topological invariant in the presence of disorder as
\begin{equation}
  \dis{Q}_{\rm bulk} = \dis Q_{\rm boundary}.
\end{equation}
Figure \ref{fig:generator} illustrates the $\ZZ_2$ group structure of disordered TCPs in class $\mbox{AIII}^{{\cal M}_{+}}$ that this identification implies.

The intrinsic boundary invariant $\dis{Q}_{\rm boundary}$ is robust to the presence of disorder if the bulk mobility gap remains open. This does not mean, however, that the addition of disorder has no effect on the configuration of zero-energy corner states in a specific crystal. The reason is that, even if the bulk gap remains open, the boundary mobility gap may close upon adding disorder. If that happens, the zero-energy corner states disappear and the system acquires delocalized states along the entire boundary. If, upon further increasing the disorder strength, the boundary mobility gap reopens, zero-energy corner states reappear. These have the same intrinsic invariant $\dis Q_{\rm boundary}$ as before the disorder-induced closing of the boundary gap, but the extrinsic invariant $\dis N_{\rm corner}$ may be different. For example, upon increasing the disorder strength, a TCP may go through a transition between having two zero-energy corner states to having none or vice versa, because both configurations of corner states correspond to the same topological invariant $\dis{Q}_{\rm boundary}$.

Without disorder, a band structure with second-order bulk topology may have accidental non-topological gapless boundary states, which obscure the topological corner states associated with the bulk topology. In the presence of disorder, the one-dimensional crystal boundaries are Anderson localized, except at fine-tuned parameters corresponding to a topological transition of the crystal boundary (see above), so that the zero-energy corner states become the {\em generic} manifestation of the bulk topology.

{\em Lattice model.---} In Sec.\ \ref{sec:scattertheory} we present lattice models of a TCP in class $\mbox{AIII}^{\mathcal{M}_+}$. Using a numerical simulation, we explicitly verify that in the presence of disorder TCPs with bulk invariant $Q_{\rm bulk} = +1$ and $Q_{\rm bulk} = -1$ can be continuously deformed into each other without closing the bulk mobility gap, whereas this is not possible for TCPs with $Q_{\rm bulk}$ differing by one. We also present a numerical example illustrating that increasing the disorder strength may induce a transition between two different corner state configurations that correspond to the same intrinsic invariant $\dis{Q}_{\rm boundary}$.

{\em Summary.---} Without disorder, the bulk topology of a class AIII$^{{\cal M}_+}$ TCP in $d=2$ manifests as zero-energy corner states with well-defined mirror parity. The bulk topology is in one-to-one correspondence with the number of corner states modulo ``extrinsic'' configurations of corner states that are associated with a decoration of the crystal boundary by one-dimensional TCPs. Disorder blurs the distinction between corner states of opposite mirror parity and, hence, trivializes certain configurations of corner states that were topologically nontrivial in the absence of disorder. Via the bulk-boundary correspondence, the ensuing simplification of the boundary classification also implies a simplification of the classification of bulk TCPs in the presence of disorder.

\subsection{Mirror-symmetric superconductor with hinge modes}
\label{sec:Ex2}

The second example is of a three-dimensional TCP --- a topological crystalline superconductor --- in class D with an additional mirror symmetry, which has hinge modes along mirror-symmetric hinges. It illustrates the appearance of a second-order statistical topological phase in the presence of disorder. For a TCP in class D, particle-hole symmetry imposes the symmetry constraint
\begin{equation}
  H(k_x,k_y,k_z) = - H(-k_x,-k_y,-k_z)^*.
  \label{eq:particleholeD}
\end{equation}
We additionally impose a mirror symmetry $z \to -z$ that commutes with particle-hole conjugation,
\begin{equation}
  H(k_x,k_y,k_z) = \tau_3 H(k_x,k_y,-k_z) \tau_3.
  \label{eq:mirrorD}
\end{equation}
We use the symbol $\mbox{D}^{{\cal M}_{+}}$ to refer to this symmetry class. The superscript ``${\cal M}_+$'' indicates the presence of a mirror symmetry that commutes with particle-hole conjugation.
As in the previous example, we first discuss the topological classification in the absence, then in the presence of disorder.

\begin{figure}
  \centering
  \includegraphics[width=0.85\columnwidth]{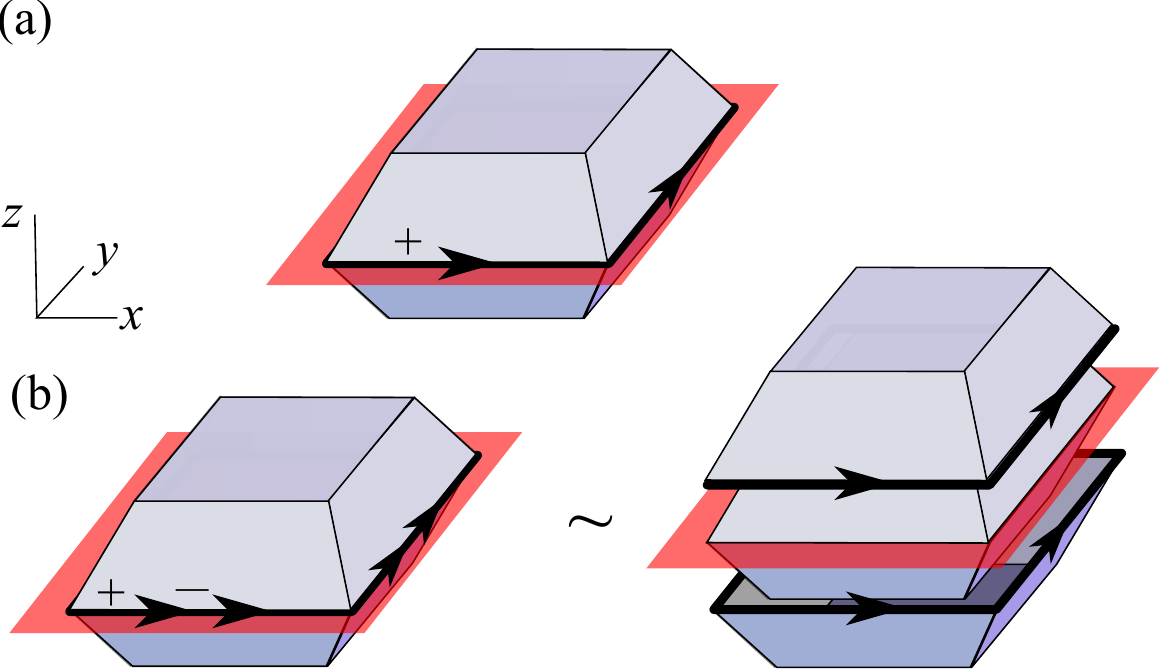}
  \caption{(a) A mirror-symmetric second-order topological superconductor in class D has chiral modes at mirror-symmetric hinges. These hinge modes have a well-defined mirror parity $\pm$. (b) A pair of co-propagating hinge modes of opposite mirror parity can be obtained by a boundary decoration. As such, the intrinsic boundary invariant counts the difference of the numbers of hinge modes of positive and negative mirror parity. \label{fig:Dstack}}
\end{figure}

{\em Non-disordered case.---} 
The bulk and boundary classifications in the absence of disorder proceed analogous to that of the previous example. At the high-symmetry planes $k_z = 0$/$\pi$, $H(k_x,k_y,k_z)$ is the diagonal sum of blocks $H_{\tau}(k_x,k_y,0/\pi)$ with mirror parity $\tau = \pm$. 
The topological invariants of the bulk band structure are obtained from the ``mirror Chern numbers'' $C_{\tau}(0)$ and $C_{\tau}(\pi)$, $\tau = \pm$, which are the Chern numbers associated with these diagonal blocks. The mirror Chern numbers satisfy the constraint $C_{+}(0) + C_{-}(0) = C_{+}(\pi) + C_{-}(\pi)$. To rule out weak topology, we further impose that $C_{+}(0) = C_{-}(0) = 0$. The remaining integer topological invariant is
\begin{equation}
  Q_{\rm bulk} = C_{-}(\pi).
\end{equation}
The anomalous boundary states are chiral Majorana modes at mirror symmetric hinges, see Fig.\ \ref{fig:Dstack}(a). The extrinsic boundary invariants are the differences
\begin{equation}
  N_{{\rm hinge},\tau} = n_{+\tau} - n_{-\tau}, \label{eq:NhingeD}
\end{equation}
where $n_{\sigma \tau}$ is the number of hinge modes of mirror parity $\tau$ propagating in direction $\sigma = \pm$. Without disorder, scattering between hinge modes of different mirror parity is not possible. Boundary decorations may simultaneously add the same integer to $N_{{\rm hinge},+}$ and $N_{{\rm hinge},-}$, see Fig.\ \ref{fig:Dstack}(b), so that the intrinsic boundary invariant is 
\begin{equation}
  Q_{\rm boundary} = N_{{\rm hinge},+} - N_{{\rm hinge},+}.
\end{equation}
The intrinsic boundary invariant is in one-to-one correspondence with the bulk invariant,
\begin{equation}
  Q_{\rm bulk} = Q_{\rm boundary}.
\end{equation}

\begin{figure}
  \centering
  \includegraphics[width=0.99\columnwidth]{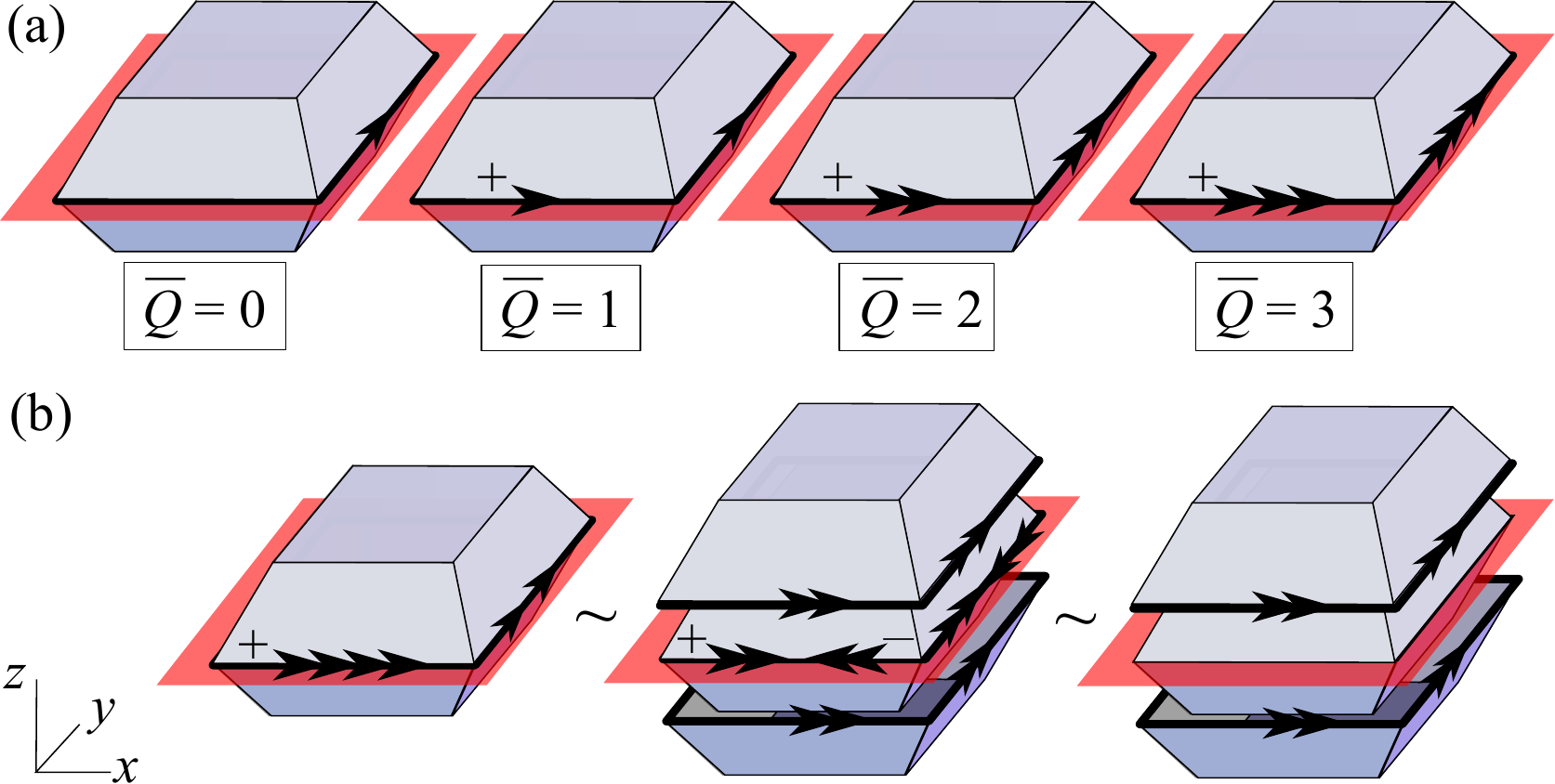}
  \caption{(a) Schematic representations of the four topological equivalence classes of disordered TCPs in class $\mbox{D}^{{\cal M}_+}$, corresponding to bulk topological invariants $\dis{Q}_{\rm bulk} = 0$, $1$, $2$, and $3$. (b) Illustration of the $\ZZ_4$ group structure of the topology, which implies that a TCP with four co-propagating modes of the same mirror parity along a mirror-symmetric hinge can be trivialized. The trivialization involves a boundary deformation to a TCP with two pairs of counter-propagating hinge modes with opposite mirror parity, which is then trivialized by the presence of disorder.
\label{fig:Dstackgroup}}
\end{figure}

{\em Disordered case.---} Although disorder may cause back-scattering between hinge modes irrespective of their mirror parity, this does not automatically lead to Anderson localization: For disorder that obeys the mirror symmetry on average, exponential localization of hinge states occurs only if $N_{{\rm hinge},+} = - N_{{\rm hinge},-}$ is even (see App.\ \ref{app:statistical} for details). As a result, in the presence of disorder, the boundary is described by one integer extrinsic invariant and one extrinsic invariant of $\ZZ_2$-type,
\begin{align}
  \dis{N}_{\rm hinge} =&\, N_{{\rm hinge},+} + N_{{\rm hinge},-},\nonumber \\
  \dis{N}'_{\rm hinge} =&\, N_{{\rm hinge},-} \mod 2.
\end{align}
The role of boundary decorations is the same with disorder that respects the mirror symmetry on average as without disorder: Boundary decorations may simultaneously change $\dis{N}_{\rm hinge}$ by two and $\dis{N}'_{\rm hinge}$ by one. This leaves the $\ZZ_4$-number
\begin{equation}
  \dis{Q}_{\rm boundary} = N_{{\rm hinge},+} - N_{{\rm hinge},-} \mod 4
\end{equation}
as the only intrinsic boundary invariant. Since there are no first-order phases, a bulk topological invariant may again be defined via the bulk-boundary correspondence,
\begin{equation}
  \dis{Q}_{\rm bulk} = \dis{Q}_{\rm boundary}.
\end{equation}
The group structure of disordered TCPs in class $\mbox{D}^{{\cal M}_+}$ is illustrated in Fig.\ \ref{fig:Dstackgroup}.

{\em Statistical second-order topology.---}
The case $\dis{Q}_{\rm bulk} = 2 \mod 4$ is an example of a ``statistical second-order superconductor''. This is a generalization of a ``statistical topological insulator'',\cite{fulga2014} a topological phase with disorder that respects a defining symmetry on average and that has boundary states that evade exponential localization. For the present example, the statistical second-order topological phase has hinge states with $N_{{\rm hinge},+} = -N_{{\rm hinge},-}$ odd (up to boundary decorations). Such counter-propagating hinge states are ``critical'' if the disorder preserves the mirror symmetry on average: Instead of exponential localization, they show power-law correlations at large distances. The same power-law correlations appear in a one-dimensional model of coupled masses and springs originally proposed by Dyson\cite{dyson1953} and in lattice models of fermions hopping on a one-dimensional chain with random nearest-neighbor hopping amplitudes.\cite{theodorou1976,eggarter1978} Correspondingly, such states do not contribute to quantized transport coefficients that are usually associated with topological phases. Nevertheless, the intrinsic invariant $\dis{Q}_{\rm bulk} = 2 \mod 4$ describes a bona fide topological phase of the bulk band structure, which is separated from band structures with different bulk invariant $\dis{Q}_{\rm bulk}$ by closings of the bulk mobility gap.

{\em Summary.---} A mirror-symmetric second-order TCP in $d=3$ has chiral hinge modes with well-defined mirror parity at a mirror-symmetric hinge. Without disorder, a pair of counter-propagating hinge modes with opposite mirror parity is protected from backscattering and, hence, a signature of a topologically nontrivial bulk. Disorder blurs the distinction between hinge modes with different mirror parity. Nevertheless, in the example considered here, the disorder-induced scattering only causes hinge modes to Anderson localize if their number is a multiple of four. If the number of hinge modes is two (modulo four), the hinge modes become critical, with power-law correlations. This is an example of a statistical second-order TCP.

\subsection{Inversion-symmetric topological insulator}
\label{sec:Ex3}

The third example is that of a three-dimensional TCP with inversion and time-reversal symmetry, so that the Bloch Hamiltonian $H(k_x,k_y,k_z)$ satisfies the symmetry constraints
\begin{align}
  H(k_x,k_y,k_z) =&\, \sigma_2 H(-k_x,-k_y,-k_z)^* \sigma_2 \nonumber \\ =&\,
  \tau_3 H(-k_x,-k_y,-k_z) \tau_3.
\end{align}
The Pauli matrices $\sigma_2$ and $\tau_3$ refer to spin and orbital degrees of freedom, respectively, whereby the eigenvalue $\tau$ of $\tau_3$ indicates the inversion parity of the orbital.
This symmetry class is denoted $\mbox{AII}^{{\cal I}_+}$, where the superscript ``${\cal I}_+$'' indicates the presence of an inversion symmetry commuting with time reversal. Because inversion symmetry is already broken at the surface, the addition of disorder does not lead to a further breaking of symmetries at the surface. Hence, disorder leaves the classification of topological phases with first- and second-order boundary states unchanged. However, as we see below, it trivializes atomic-limit phases without a filling anomaly.

{\em Non-disordered case.---} To calculate the bulk topological invariant, we note that all bands are twofold degenerate because of Kramers' theorem. The bulk invariant is computed from the differences $D_{\tau}(S) = n_{\tau}(S) - n_{\rm occ}$ at the eight high-symmetry momenta $S = (S_x,S_y,S_z)$ with $S_x = S_y = S_z = 0 \mod \pi$, where $n_{\tau}(S)$ is the number of occupied Kramers-degenerate pairs of bands with inversion parity $\tau$ at $S$ and $4 n_{\rm occ}$ is the total number of filled bands. These numbers satisfy the constraint $D_{+}(S) + D_{-}(S) = 0$ for all $S$. We consider band structures without weak invariants and impose $D_{+}(S) = D_{-}(S) = 0$ for all high-symmetry points except $S = (\pi,\pi,\pi)$. The remaining bulk topological invariant that indicates a strong phase is then the integer
\begin{equation}
  Q_{\rm bulk} = D_{-}(\pi,\pi,\pi).
\end{equation}
Band structures with $Q_{\rm bulk} = 1 \mod 4$ or $Q_{\rm bulk} = 3 \mod 4$ are first-order topological insulators with Dirac-cone surface states on all crystal surfaces.\cite{fu2007b} Band structures with $Q_{\rm bulk} = 2 \mod 4$ have helical hinge modes,\cite{khalaf2018,khalaf2018b} whereas band structures with $Q_{\rm bulk} = 0 \mod 4$ are of atomic-limit type. 

\begin{figure}
\centering
	\includegraphics[width=0.4\columnwidth]{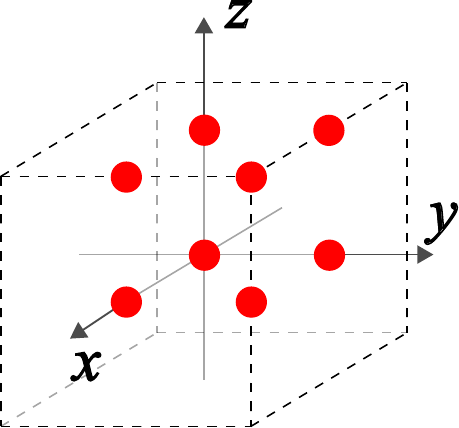}
	\caption{\label{fig:Wyckoff3d} Special Wyckoff positions for a three-dimensional inversion symmetric crystal. The eight special Wyckoff positions are at $\vr_{\rm w} = (x_{\rm w},y_{\rm w},z_{\rm w})$ with $x_{\rm w}$, $y_{\rm w}$, $z_{\rm w} \in \{0,\tfrac{1}{2}\}$, with the unit cell's size set to one and the origin at its center.}
\end{figure}

Following Ref.\ \onlinecite{vanmiert2018}, atomic-limit band structures may also be obtained from a real-space picture. Hereto, we build atomic-limit phases from Kramers-degenerate orbitals of even ($+$) and odd ($-$) inversion parity localized at one of the eight special Wyckoff positions $\vr_{\rm w} = (x_{\rm w},y_{\rm w},z_{\rm w})$ with $x_{\rm w}$, $y_{\rm w}$, $z_{\rm w} \in \{0,1/2\}$, see Fig.\ \ref{fig:Wyckoff3d}. We denote the difference of the numbers of occupied Kramers-degenerate orbitals of even and odd inversion parity $\tau$ at the Wyckoff position w by $\Delta n(\vr_{\rm w})$. The parity of the numbers $\Delta n(\vr_{\rm w})$ are topological invariants and satisfy the parity constraint
\begin{equation}
  \sum_{{\rm w}} \Delta n(\vr_{\rm w}) = 0 \mod 2.
\end{equation}
The topological numbers $D_{\tau}(S)$ can be expressed in terms of the eight real-space invariants $\Delta n(\vr_{\rm w})$,
\begin{equation}
  D_{\tau}(S) = \frac{1}{2} \sum_{{\rm w}} e^{2 i \vk_{S} \cdot \vr_{\rm w}}
  \Delta n(\vr_{\rm w}).
\end{equation}
In the absence of weak invariants, atomic limits satisfy
\begin{equation}
  Q_{\rm bulk} = 4 \Delta n(1/2,1/2,1/2).
  \label{eq:Qbulkreal}
\end{equation}

Atomic-limit insulators with $Q_{\rm bulk} = 4 \mod 8$ have a {\em filling anomaly}:\cite{benalcazar2019} A crystal that is globally inversion symmetric and that has a charge-neutral unit cell, but where the parity of the entire system's net charge is odd. This remains true if translation symmetry is broken and a superlattice with a larger unit cell is formed, provided the global inversion symmetry of the lattice remains in place. The additional charge is generically localized at two inversion related corners, each harboring a half-integer charge.

Conversely, for an inversion-symmetric crystal one defines intrinsic boundary invariants $Q_{{\rm boundary},1}$, $Q_{{\rm boundary},2}$, $F \in \{0,1\}$ signaling the presence or absence of anomalous first- and second-order boundary features or a filling anomaly, respectively.
\footnote{For band structures without surface states, one defines $Q_{{\rm boundary},2}$ as the number of helical hinge modes crossing a generic plane $\Omega$ through the inversion center $\mod 4$. For band structures with anomalous surface states, one takes the direct sum with a reference band structure with $Q_{\rm bulk} = 7 \mod 8$, adds a surface perturbation that gaps out the surface states, and then determines $Q_{{\rm boundary},2}$ as described above. In the same, way, one may establish the presence or absence of a filling anomaly for band structures with anomalous hinge or surface states.} 
One then has the bulk-boundary correspondence
\begin{equation}
  Q_{{\rm boundary},1} + 2 Q_{{\rm boundary},2} + 4 F = Q_{\rm bulk} \mod 8.
\end{equation}
This bulk-boundary correspondence is incomplete: $Q_{\rm bulk}$ determines the boundary invariants $Q_{{\rm boundary},1}$, $Q_{{\rm boundary},2}$, and $F$, but the boundary invariants determine the integer $Q_{\rm bulk}$ only up to a multiple of eight. This reflects the fact that atomic-limit band structures cannot be distinguished via the presence of a anomalous boundary states and that not all atomic-limit band structures have a filling anomaly.

{\em Disordered case.---} Even without disorder, inversion symmetry is already broken everywhere at the crystal boundary. This is a key difference compared to the case of mirror symmetry discussed in the previous two examples, which is preserved at mirror-symmetric corners and hinges. Hence, the presence of disorder locally does not lead to any further breaking of symmetries. It follows that the boundary invariants $\dis{Q}_{{\rm boundary},1}$ and $\dis{Q}_{{\rm boundary},2}$ in the presence of disorder are precisely the same as their counterparts in the non-disordered case,
\begin{align}
  \dis{Q}_{{\rm boundary},1} =&\, Q_{{\rm boundary},1}, \nonumber \\
  \dis{Q}_{{\rm boundary},2} =&\, Q_{{\rm boundary},2}.
\end{align}
Further, since the charge of the insulating ground state cannot change under continuous deformations, disorder that preserves the inversion symmetry on average must also preserve the filling anomaly,
\begin{align}
  \dis{F} =&\, F.
\end{align}

To obtain the bulk classification in the presence of disorder, we note that disorder trivializes all bulk phases without boundary signatures, {\em i.e.}, those with $Q_{\rm bulk} = 0 \mod 8$. This can most easily be seen in the real-space picture: Disorder blurs the distinction between even- and odd-parity orbitals at Wyckoff position $\vr_{\rm w}$, so that the real-space invariants $\Delta n(\vr_{\rm w})$ are defined up to a multiple of two only. It then follows from Eq.\ (\ref{eq:Qbulkreal}) that the bulk invariant $Q_{\rm bulk}$ is defined up to a multiple of $8$,\footnote{That the bulk phase with $Q_{\rm bulk} = 8$ is trivialized by disorder can be seen explicitly by constructing an insulator with $Q_{\rm bulk} = 8$ by placing two odd-parity Kramers doublets at Wyckoff positions $(0,0,0)$, $(0,1,1)$, $(1,0,1)$, and $(1,1,0)$ and even odd-parity Kramers doublets at Wyckoff positions $(0,0,1)$, $(0,1,0)$, $(1,0,0)$, and $(1,1,1)$. In the presence of disorder, this system may be deformed to an insulator with one Kramers doublet of each parity at each Wyckoff position, which is topologically trivial.} from which triviality of the phases with $Q_{\rm bulk} = 0 \mod 8$ immediately follows. We conclude that the bulk topological invariant in the presence of disorder is of $\ZZ_8$ type,
\begin{equation}
  \dis{Q}_{\rm bulk} = \dis{Q}_{{\rm boundary},1} 
  + 2 \dis{Q}_{{\rm boundary},2} 
  + 4 \dis{F}
  \mod 8.
\end{equation}
Since this invariant is uniquely determined by boundary invariants, the bulk-boundary correspondence is restored in the presence of disorder.

{\em Summary.---} The inversion-symmetric topological insulator is an example of a TCP that not only has higher-order topological phases, but also obstructed atomic-limit phases. The addition of disorder does not affect the first-order and second-order boundary states, but it trivializes obstructed atomic-limit phases without a filling anomaly. As a result, for inversion-symmetric TCPs, the bulk-boundary correspondence, which was incomplete in the clean limit, is restored by the presence of disorder.

\section{Classification tables}
\label{sec:classification}

\begin{figure}
    \centering
    \includegraphics[width=0.85\columnwidth]{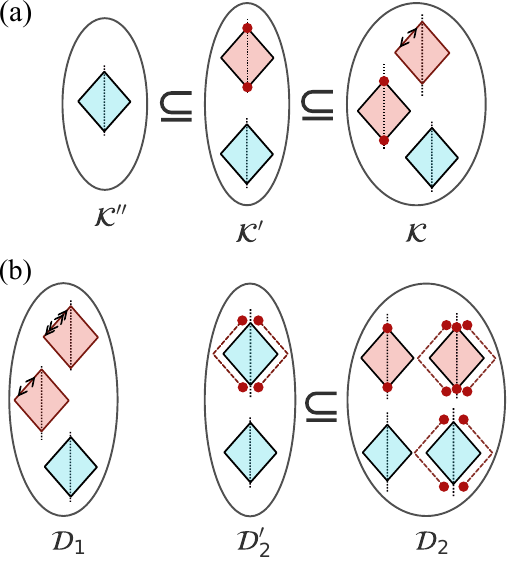}
    \caption{\label{fig:subgroups} Schematic illustration of the subgroup sequences classifying TCPs in two dimensions. 
    (a) For the bulk subgroup sequence of a two-dimensional TCP, the classifying group ${\cal K}$ classifies all bulk phases, irrespective of their boundary signature, whereas ${\cal K}'$ and ${\cal K}''$ are subgroups classifying bulk TCPs without first-order, and without first- or second-order boundary states, respectively. A red/blue bulk indicates the presence/absence of intrinsic boundary states. 
    (b) The boundary classification of a two-dimensional TCP: The group ${\cal D}_1$ classifies all anomalous first-order boundary states; the subgroup sequence ${\cal D}'_{ 2} \subseteq {\cal D}_{ 2}$ involves the classifying group ${\cal D}_2$ of all zero-energy states at a high-symmetry corner and the subgroup ${\cal D}_2'$ classifying zero-energy states that can be obtained by adding a boundary decoration to a trivial bulk.}
\end{figure}

\begin{table*}
\begin{tabular*}{\linewidth}{c@{\extracolsep{\fill}} cc  cc }
 \hline\hline
 Class & 
 \multicolumn{2}{ c }{ Bulk classification} &
 \multicolumn{2}{ c }{ 2\textsuperscript{nd} order boundary }\\
 &
 $\mathcal{K}'' \!\subseteq\! \mathcal{K}' \!\subseteq\! \mathcal{K}$&
 $\dis{\mathcal{K}}'' \!\subseteq\! \dis{\mathcal{K}}' \!\subseteq\! \dis{\mathcal{K}}$ &
 $\mathcal{D}_2' \!\subseteq\! \mathcal{D}_2$ &
 $\dis{\mathcal{D}}_2' \!\subseteq\! \dis{\mathcal{D}}_2$\\
 \hline
 $\text{AIII}^\mathcal{M_+}$    & $ 0 \subseteq \ZZ   \subseteq \ZZ     $ & $ 0 \subseteq \ZZ_2 \subseteq \ZZ_2 $ & $ \ZZ    \subseteq \ZZ^2 $ & $ 2\ZZ \subseteq \ZZ  $ \\
 \hline\hline
\end{tabular*}
\caption{Subgroup sequences for symmetry class $\mbox{AIII}^{{\cal M}_+}$, $d=2$, summarizing the discussion of Sec. \ref{sec:Ex1}. To relate the entries in the table to the discussion of Sec.\ \ref{sec:Ex1}, note that two integers $N_{{\rm corner},\pm}$ are required to label anomalous configurations of zero-energy corner states, see Eq.\ (\ref{eq:NcornerAIII}). This explains the extrinsic classifying group ${\cal D}_{ 2} = \ZZ^2$. Since configurations of corner states with $N_{{\rm corner},+} = N_{{\rm corner},-}$ can be obtained from boundary decorations of a trivial insulator, ${\cal D}_{ 2}' = \ZZ$ is the diagonal subgroup of ${\cal D}_{ 2}$.
\label{tab:casestudy1}}
\end{table*}

\begin{table*}
\begin{tabular*}{\linewidth}{c@{\extracolsep{\fill}} cc  cc cc }
 \hline\hline
 Class & 
 \multicolumn{2}{ c }{ Bulk classification} &
 \multicolumn{2}{ c }{ 3\textsuperscript{rd} order boundary } &
 \multicolumn{2}{ c }{ 2\textsuperscript{nd} order boundary }\\
 &
 $\mathcal{K}''' \!\subseteq\!\mathcal{K}'' \!\subseteq\! \mathcal{K}' \!\subseteq\! \mathcal{K}$&
 $\dis{\mathcal{K}}''' \!\subseteq\! \dis{\mathcal{K}}'' \!\subseteq\! \dis{\mathcal{K}}' \!\subseteq\! \dis{\mathcal{K}}$ &
 $\mathcal{D}_3'' \!\subseteq\! \mathcal{D}_3' \!\subseteq\! \mathcal{D}_3$ &
 $\dis{\mathcal{D}}_3'' \!\subseteq\! \dis{\mathcal{D}}_3' \!\subseteq\! \dis{\mathcal{D}}_3$ &
 $\mathcal{D}_2' \!\subseteq\! \mathcal{D}_2$ &
 $\dis{\mathcal{D}}_2' \!\subseteq\! \dis{\mathcal{D}}_2$\\
 \hline
 $\text{D}^\mathcal{M_+}$     & $ 0    \subseteq 0    \subseteq \ZZ   \subseteq \ZZ $ & $ 0     \subseteq 0     \subseteq \ZZ_4 \subseteq \ZZ_4 $ & $ \ZZ_2^2 \subseteq \ZZ_2^2 \subseteq \ZZ_2^2 $ & $ \ZZ_2 \subseteq \ZZ_2 \subseteq \ZZ_2  $ & $ \ZZ   \subseteq \ZZ^2 $ & $\bl{2\ZZ\subseteq\ZZ\times\ZZ_2^*}$\\
 $\text{AII}^{\mathcal{I}_+}$ & $ 4\ZZ \subseteq 4\ZZ \subseteq 2\ZZ  \subseteq \ZZ $ & $ \ZZ_2 \subseteq \ZZ_2 \subseteq \ZZ_4 \subseteq \ZZ_8 $ & $ 0       \subseteq 0       \subseteq 0       $ & $ 0     \subseteq 0     \subseteq 0      $ & $ 0  \subseteq \ZZ_2  $ & $0 \subseteq \ZZ_2 $ \\
 \hline \hline
\end{tabular*}
\caption{Subgroup sequences for symmetry classes $\text{D}^\mathcal{M_+}$ and $\text{AII}^{\mathcal{I}_+}$, $d=3$, see Secs.\ \ref{sec:Ex2} and \ref{sec:Ex3}, respectively. For class $\mbox{D}^{\mathcal{M}_+}$, the extrinsic boundary classifying group ${\cal D}_{3}'' = \ZZ_2^2$ corresponds to decorations of mirror-symmetric crystal hinges with one-dimensional TCPs. The corresponding corner states are Majorana zero modes with well-defined mirror parity. In the presence of disorder, the distinction between mirror parities is lifted, so that $\dis{\cal D}_{3}'' = \ZZ_2$. The extrinsic boundary classifying group ${\cal D}_2 = \ZZ^2$ represents the two integers $N_{{\rm hinge},\pm}$ labeling the anomalous hinge states, see Eq.\ (\ref{eq:NhingeD}), whereas the diagonal subgroup ${\cal D}_{2}' = \ZZ$ describes configurations of hinge modes that correspond to surface decorations of a trivial superconductor. The subgroup relation $2 \ZZ \subseteq \ZZ \times \ZZ_2^*$ for the disordered boundary classifying groups corresponds to the embedding $2n \to (2n,n \mod 2)$, $n \in \ZZ$ (the superscript $*$ indicates a statistical higher-order phase). Hence, the classification of 2nd order phases is given by the quotient ${\cal K}'/{\cal K}'' = {\cal D}_{2}/{\cal D}_2' = \ZZ_4$, both from the bulk and from the boundary perspective. For class $\mbox{AII}^{{\cal I}_{-}}$, the subgroup ${\cal K}''' = 4 \ZZ$ labels atomic-limit phases, whereby elements $4n \in {\cal K}'''$ with $n = 1 \mod 2$ indicate obstructed atomic-limit phases with a filling anomaly.
\label{tab:casestudy2}}
\end{table*}

The arguments presented for TCPs in symmetry classes $\mbox{AIII}^{{\cal M}_+}$ in $d=2$ and for $\mbox{D}^{{\cal M}_+}$ and $\mbox{AII}^{{\cal I}_+}$ in $d=3$ in Sec.\ \ref{sec:casestudy} can be generalized to all other symmetry classes and dimensions. Following Ref.\ \onlinecite{trifunovic2019}, we present the results of such a generalization in terms of {\em subgroup sequences} for bulk and boundary classifications, which summarize the classification for each symmetry class. 

The bulk classification is represented by the subgroup sequence
\begin{equation}
  {\cal K}''' \subseteq {\cal K}'' \subseteq {\cal K}' \subseteq {\cal K},
\end{equation}
where ${\cal K}$ is the full classifying group, ${\cal K}' \subseteq {\cal K}$ the subgroup containing topological phases without first-order boundary states, ${\cal K}'' \subseteq {\cal K}'$ the subgroup of ${\cal K}'$ containing topological phases without first- or second-order boundary states, and ${\cal K}'''$ the subgroup containing phases without any boundary states, {\em i.e.} obstructed atomic limits. (For two dimensional TCPs, ${\cal K}'''$ is omitted.) 

For the boundary classification we introduce the classifying group ${\cal D}_{ 1}$ of anomalous first-order boundary states and subgroup sequences for anomalous second-order and third-order boundary states,
\begin{align}
  {\cal D}'_{ 2} \subseteq&\, {\cal D}_{ 2},  \nonumber \\
  {\cal D}''_{3} \subseteq {\cal D}'_{3} \subseteq&\, {\cal D}_3.
  \label{eq:subboundary}
\end{align}
Here ${\cal D}_{2}$ and ${\cal D}_{3}$ classify all configurations of anomalous second-order and third-order boundary states, respectively. This includes both intrinsic as well as extrinsic higher-order boundary states that can be obtained by ``decorating the boundary'' of a crystal with lower-dimensional TCPs. 
Furthermore, ${\cal D}'_{ 2}$ and ${\cal D}'_{3}$ are the subgroups classifying all configurations of second-order and third-order boundary states, respectively, that can be obtained from boundary decoration on the surface of a trivial insulating bulk. 
The subgroup ${\cal D}_{3}''$ (for three-dimensional TCPs) classifies configurations of third-order boundary states that can be obtained by means of a decoration on a trivial insulator that only has support on the crystal hinges. 
Both subgroup sequences are illustrated schematically for a two-dimensional TCP in Fig.\ \ref{fig:subgroups}. 

With this notation, first-, second-, and third-order TCPs are classified by the quotients ${\cal K}/{\cal K}'$, ${\cal K}'/{\cal K}''$, and ${\cal K}''/{\cal K}'''$, respectively. Similarly, the intrinsic boundary classifications for first-, second-, and third-order boundary states are ${\cal D}_{ 1}$, ${\cal D}_{ 2}/{\cal D}_{ 2}'$, and ${\cal D}_{3}/{\cal D}_{3}'$, respectively. The bulk-boundary correspondences then read
\begin{align}
  {\cal K}/{\cal K}'     =&\, {\cal D}_{ 1}               ,\nonumber \\
  {\cal K}'/{\cal K}''   =&\, {\cal D}_{ 2}/{\cal D}_{ 2}',\nonumber \\
  {\cal K}''/{\cal K}''' =&\, {\cal D}_{ 3}/{\cal D}_{ 3}',
\end{align}
where the third equality only applies to three-dimensional TCPs. For TCPs in $d=1$, $d=2$ and $d=3$ the bulk-boundary correspondence is complete if ${\cal K}' = 0$, ${\cal K}'' = 0$ and ${\cal K}''' = 0$, respectively, or if equal to $\ZZ_2$ and the nontrivial phase has a filling anomaly. 

To illustrate these definitions, we summarize the results for the three symmetry classes discussed in Sec.\ \ref{sec:casestudy} in Tabs.\ \ref{tab:casestudy1} and \ref{tab:casestudy2}. Classifying groups for disordered TCPs are denoted with an overline $\dis{\cdots}$. The figure captions contain a few additional remarks indicating how the entries in the tables relate to the discussion in Sec.\ \ref{sec:casestudy}.

Complete classification tables similar to Tabs.\ \ref{tab:casestudy1} and \ref{tab:casestudy2} for all other TCPs with mirror, twofold rotation, or inversion symmetry in $d=1$, $d=2$ and $d=3$ are given in App.\ \ref{app:classification}. These tables generalize the boundary-resolved classifications for the non-disordered case that are available in the literature.\cite{geier2018,khalaf2018b,trifunovic2019}

Our classification results for mirror-symmetric TCPs with $d=2$ partially overlap with the classification obtained in Ref.\ \onlinecite{spring_amorphous_2021} for statistical topological insulators in two-dimensional amorphous systems with an average (unitary) mirror symmetry and a continuous rotation symmetry. There are differences for symmetry classes $\text{DIII}^{\mathcal{M}_{--}}$, $\text{DIII}^{\mathcal{M}_{-+}}$ and $\text{CII}^{\mathcal{M}_{--}}$, which are trivial in the classification of Ref.\ \onlinecite{spring_amorphous_2021}, whereas our boundary-based classification gives a $\mathbb{Z}_2$ invariant with anomalous zero-energy states at mirror-symmetric corners. We attribute the difference to the additional continuous rotation symmetry in Ref.\ \onlinecite{spring_amorphous_2021}.

\section{Microscopic Models}
\label{sec:microscopics}

In this section we show how the phenomenology described in Sec.\ \ref{sec:casestudy} manifests in  microscopic models. We discuss two case studies. In the first, we study a second-order topological insulator in three dimensions with inversion symmetry (class $\text{A}^{\mathcal{I}}$). We show, by way of a field theoretic approach, that hinge modes in such a system are robust to disorder preserving inversion symmetry on average and weak enough to not close surface gaps.

The second example is a study of a mirror-symmetric second-order topological insulator in two dimensions with chiral symmetry (class $\text{AIII}^{\mathcal{M_+}}$). Using transport simulations with the kwant package,\cite{groth2014} we show that in the presence of disorder, lattice Hamiltonians with different non-disordered (boundary) invariants $Q_{\rm boundary}$, but the same disordered invariant $\overline{Q}_{\rm boundary}$ can be deformed into one another without closing the bulk mobility gap, whereas this is not possible if the disordered invariants $\overline{Q}_{\rm boundary}$ are different. This illustrates the arguments of Sec\ \ref{sec:Ex1} that in the presence of disorder there is a bulk-boundary correspondence, so that the boundary invariant $\overline{Q}_{\rm boundary}$ also describes the bulk topology. We also show (for a different model in the same symmetry class) that increasing the strength of the disorder alone can lead to a closing and subsequent reopening of the boundary gap and can, thus, induce a transition between different extrinsic boundary signatures, while leaving the intrinsic topological invariant $\overline{Q}_{\rm boundary}$ unchanged.

\subsection{Inversion-symmetric insulator with hinge modes}
\label{sec:fieldtheory}

\begin{figure}
    \centering
    \includegraphics[width = 0.74\columnwidth]{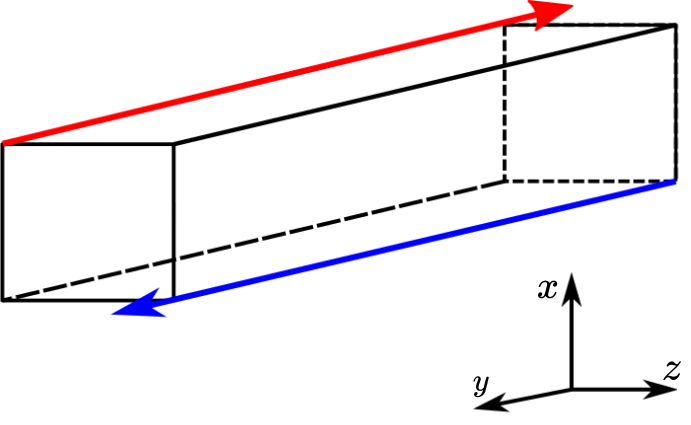}
    \caption{The model (\ref{axi}) in a parallelepiped geometry. In this geometry the model can host chiral modes propagating in two of the hinges.}
    \label{fig:inversion}
\end{figure}

As our first example, we consider a three-dimensional second-order TCP in class A$^\mathcal{I}$. Inversion symmetry is implemented as $\mathcal{I} = \tau_1$, and acts on the Hamiltonian as,
\begin{align}
  H(k_x,k_y,k_z) = \tau_1 H(-k_x,-k_y,-k_z) \tau_1.
\end{align}
We take, as a concrete model, the Hamiltonian
\begin{align}
\begin{split}
    H (k) =& \left( M-\sum_{i=1}^3 \cos{k_i} \right) \tau_1\sigma_0 \\ 
    &\, + v \sum_{i=1}^3 \sin{k_i} \tau_3 \sigma_i + m \sum_{i=1}^3 \tau_0 \sigma_i,
    \label{axi}
\end{split}
\end{align}
where $\tau_a$ and $\sigma_a$ ($a=0,1,2,3$) are two sets of Pauli matrices plus the identity, $M$ is the bulk mass, $m \ll 1$ is a small parameter that gaps out surface states along coordinate planes, and $v$ is the velocity. A topological phase that  supports two chiral hinge modes occurs when $1<|M|<3$. These are located at inversion related hinges and propagate in opposite directions, see Fig.\ \ref{fig:inversion}.

We construct an effective surface Hamiltonian for a system that is infinitely long in the $y$-direction but finite in the $x$ and $z$-directions. To this end, we take Eq.\ (\ref{axi}) in the topological phase and project onto zero-energy surface states. This yields an effective low energy Hamiltonian for each of the 4 surfaces. These can be combined into a single surface theory describing a plane that is infinite in $y$ but periodic in a second coordinate $x'$ (topologically a cylinder). The resulting $2d$ effective surface Hamiltonian takes the form\cite{schindler2018, benalcazar2017}
\begin{equation}
    H_0 = v k_{x'} \sigma_1 + v k_y \sigma_2 + m(x')\sigma_3, 
    \label{eq:H0eff}
\end{equation}
where the parameter $m$ acquires a spatial dependence. On half of the surface $m>0$ while the other half $m<0$, thereby resulting in chiral hinge modes, each with support on one of the two hinges where $m$ changes sign. The modulus $|m|$ is the surface spectral gap. The effective Hamiltonian (\ref{eq:H0eff}) is valid so long as the surface gap $|m|$ is much smaller than the bulk gap of the full three-dimensional lattice Hamiltonian (\ref{axi}).

We next evaluate the effect of surface disorder, in particular, we would like to see the stability of the hinge modes under the influence of random disorder. We take the effective surface Hamiltonian  (\ref{eq:H0eff}) and model the disorder by a scalar Gaussian distributed potential $V(\mathbf{x})$, 
\begin{align}
    \begin{split}
         \left< V(\mathbf{x}) \right>_{\rm dis} &= 0, \\
    \left< V(\mathbf{x}) V(\mathbf{x}_0) \right>_{\rm dis} &= \frac{\gamma_0}{2}\delta( \mathbf{x}-\mathbf{x}_0).
    \end{split}
\end{align}
with $\mathbf{x} \equiv (x',y)$.

The surface Hamiltonian of the disordered system takes the form 
\begin{equation}
  H_S = H_0 + V(\mathbf{x})\sigma_0. \label{eq:HS}
\end{equation}

We treat the disordered problem by way of a replica field theory.\cite{altland2010condensed} After disorder averaging, and as an intermediate result, we obtain an action depending on the Goldstone mode $Q=T^{-1}\tau_3 T$, $T \in U(2R)/(U(R) \times U(R))$ and an external gauge field $\mathbf{a}$,
\begin{align}
 \begin{split}
 S[Q,\mathbf{a}] =&\, S_0[Q,\mathbf{a}]- S^{M}_{\eta}[Q,\mathbf{a}] \\ 
 =&\,         -\operatorname{tr} \ln{(\epsilon + v(\mathbf{a} \cdot \mathbf{\sigma}) - v(\mathbf{k} \cdot \mathbf{\sigma}) - m\sigma_3 + i \kappa Q)}\\  
  &\, \mbox{} +\operatorname{tr} \ln{(           v(\mathbf{a} \cdot \mathbf{\sigma}) - v(\mathbf{k} \cdot \mathbf{\sigma}) - \widetilde{M}\sigma_3 + i \eta Q)},  
 \label{regactio}    
 \end{split}
\end{align}
where we have used a Pauli-Villars regularization \cite{pauli1949invariant} in order to avoid UV divergences, with $\widetilde{M}\to \infty$ and $\eta \to 0$, and $\kappa$ is the scattering rate off impurities, which is found self-consistently as, 
\begin{equation}
    \kappa = \gamma_0 \operatorname{Im} \left[ \int \frac{d^2 k}{(2\pi)^2} \operatorname{Tr}\left(\frac{1}{\epsilon-i\kappa-H_0} \right) \right].
\end{equation}
We expand the action (\ref{regactio}) in gradients of $T$ to construct a low energy action that takes the form of a nonlinear $\sigma$ model,\cite{ostrovsky2007,koenig2014halfinteger}
\begin{align}
 \begin{split}
    S[Q] =& \frac{1}{8} \Big( \sigma_{xx} \int d^2 x \operatorname{tr}(\nabla Q)^2 \\ & \, + \sigma_{xy}\int d^2 x \varepsilon^{ij3} \operatorname{tr}(Q\nabla_iQ\nabla_jQ)\\ \, & \,-\frac{1}{2} \int d^2 x \varepsilon^{ij3} \operatorname{tr}(Q\partial_iQ\partial_jQ) \Big),  \label{nlsm1}
 \end{split}
\end{align}
where $Q = T^{-1}\tau_3 T$, $\nabla_i = \partial_i - i[a_i, \quad]$ is the
covariant derivative, 
\begin{align*}
    \sigma_{xx} &= \frac{1}{2\pi} \left(1+\frac{\epsilon^2+\kappa^2-m^2}{2\kappa} f(\epsilon,m) \right), \\
    \sigma_{xy} &= \frac{m}{2\pi}(f(\epsilon,m)+f(m,\epsilon)),
\end{align*}
with 
\begin{align*}
    f(x,y) =
\frac{1}{x} \left( \arctan \left( \frac{x+y}{\kappa} \right) + \arctan
\left(\frac{x-y}{\kappa} \right) \right),
\end{align*}
and we have set $v=1$. 

The $\theta$ angle of the nonlinear sigma model is therefore given by $\theta = 2\pi \left( \sigma_{xy}- \tfrac{1}{2} \right) \,\mathrm{mod} \,2\pi$. In the large-system-size limit, this parameter renormalizes to an integer multiple of $2\pi$,\cite{pruisken1984, levine1984theory, khmel1983quantization} $\theta \to 2\pi\, \lfloor \sigma_{xy}-\tfrac{1}{2}\rceil$, where $\lfloor . \rceil$ is the nearest integer function ($\lfloor2.6\rceil=3$, $\lfloor 1.2\rceil=1$, etc). This implies that $\sigma_{xy}$ asymptotically assumes a half-integer quantized value --- the half-integer quantum Hall effect occurring at topological insulator surfaces.  

Our model has a space-dependent mass-like parameter $m(x')$, with $m(x')>0$ in one half of the space, $m(x')<0$ the other, and a smooth interpolation in-between. The scaling behavior outlined above then implies that  $\sigma_{xy}= \tfrac{1}{2} \,\mathrm{sgn}(m)$ in the two half regions, asymptotically. The quasi-one dimensional interface region  supports one propagating quantum Hall edge channel, corresponding to the hinge mode in the three-dimensional crystal described by Eq.\ (\ref{axi}). 

\begin{figure}
    \includegraphics[width=0.98\columnwidth]{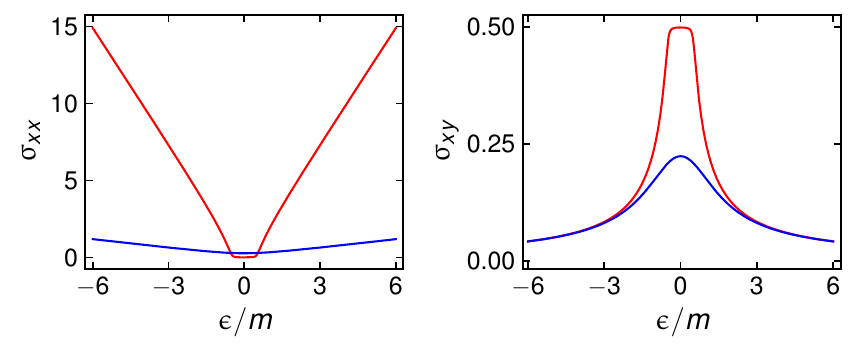}
	\caption{Longitudinal and transverse conductivities $\sigma_{xx}$ and $\sigma_{xy}$ for weak (red, $\kappa/m = 0.2$) and strong (blue, $\kappa/m = 2.6$) disorder. The surface gap is set to $m = 0.5$. Without disorder, the surface spectrum is gapped for $|\varepsilon| < |m|$. For $m<0$, $\sigma_{xy}$ changes sign, while $\sigma_{xx}$ remains the same.}
	\label{fig:conduc}
\end{figure}

How does this scenario respond to increases in the disorder strength, from values $\kappa<|m|$ to $\kappa>|m|$? In Fig.~\ref{fig:conduc}, the two `bare' transport coefficients $\sigma_{ij}$ corresponding to these regimes are shown as functions of the energy $\epsilon$, in red and blue respectively. These coefficients define the short-distance starting values for the renormalization described above. For weak disorder, the longitudinal conductance $\sigma_{xx}$ at increasing distance scales stays low inside the gap $|\epsilon|<|m|$ but grows quickly if $|\epsilon|$ exceeds $|m|$ due to the spectral weight present outside the gap. At the same time, $\sigma_{xy}$ diminishes with $|\epsilon|$, so that the bare value of $\theta$ is close to $-\frac{1}{2}$. 

For strong disorder and intra-gap energies, $\sigma_{xx}$ exceeds its weak-disorder limit due to impurity states smearing the gap. Outside the gap we observe qualitatively similar behavior as in the weak-disorder case, but with the high conductance regime is being reached much slower than in the clean case. 

Despite the very different bare values for the weak and strong disorder regimes the surfaces will, in both cases, approach an insulating state in the thermodynamic limit upon taking into account the renormalization described by the nonlinear sigma model Eq.\ (\ref{nlsm1}). This means that, asymptotically, the existence of hinge states is ensured for all energies inside the bulk gap (including energies larger than the surface spectral gap $|m|$ in the absence of disorder). The surface states may, however, have a large localization length for large energies or for weak disorder. As such, finite system sizes would compromise the hinge states because of backscattering via surface intermediaries in the regime where the surface localization length exceeds the system size.

The model Eq.\ (\ref{axi}) corresponds to an intrinsic second-order topological insulator: The presence of chiral hinge states is ensured by the nontrivial topology of the bulk. Formally, the addition of a Pauli-Villars regulator to the surface theory breaks inversion symmetry and it could be argued that the regularized model corresponds to an extrinsic second-order topological insulator. In both cases, the boundary phenomenology is the same, since the breaking of crystalline symmetry by the Pauli-Villars regulator does not close the bulk or the surface gaps. For the same reason, the (quantized) response to external sources and the fate of the hinge states upon the addition of disorder is the same in the intrinsic and extrinsic scenarios.
    
\subsection{Mirror-symmetric insulator with corner states}
\label{sec:scattertheory}

As a second example, we consider a two-dimensional second-order TCP in class $\text{AIII}^{\mathcal{M}_+}$. The classification of this symmetry class without and with disorder was already discussed in Sec.\ \ref{sec:Ex1}. Here we numerically analyze a lattice model and show that two clean realizations with different bulk topology can (cannot) be continuously deformed into each other in the presence of disorder if they have the same (different) invariants $\overline{Q}_{\rm boundary}$.

The eight-band lattice model is described by 
\begin{equation} 
  H = \begin{pmatrix} h_+ + \sigma_2 \tau_1 & \delta \sigma_1 \tau_3 \sin k_y  \\
    \delta \sigma_1 \tau_3 \sin k_y &
    h_- + \sigma_2(\tau_1 \cos k_y + \tau_2 \sin k_y ) \end{pmatrix},
    \label{eq:Hlattice}
\end{equation}
where $h_{\pm} = (M + \cos k_x) \sigma_2 \tau_0 \pm \sigma_1 \tau_0 \sin k_x $, whereby 
$M$ sets the (bulk) topological phase in the absence of disorder. The lattice model of Eq.\ (\ref{eq:Hlattice}) satisfies the chiral symmetry $H(k_x,k_y) = - \sigma_3 H(k_x,k_y) \sigma_3$ and the mirror symmetry $H(k_x,k_y) = \tau_1 H(k_x,-k_y) \tau_1$. For $\delta=0$ the model Eq.\ (\ref{eq:Hlattice}) corresponds to the layer representation of a second-order TCP in class $\text{AIII}^{\mathcal{M}_{+}}$, see App.\ \ref{app:BB} for details. In that limit $H$ is block-diagonal and the two diagonal blocks represent the two layers, which have opposite topological invariants. (Note that there is a subtle difference with App.\ \ref{app:BB}: The layer representation of App.\ \ref{app:BB} has a different choice of unit cell. For the unit cell choice of App.\ \ref{app:BB}, $H$ is manifestly without $k_y$-dependence, but the mirror symmetry is $k_y$-dependent because it mixes unit cells. The choice of unit cell corresponding to Eq.\ (\ref{eq:Hlattice}) has a $k_y$-dependent Hamiltonian and a $k_y$-independent mirror symmetry.) 

As discussed in Sec.\ \ref{sec:Ex1}, the bulk invariant in the absence of disorder may be obtained from the winding numbers $W_{\tau}(k_y) \in \ZZ$, where $k_y = 0$ or $\pi$ and the subscript $\tau$ indicates the mirror parity.\cite{chiu2013, morimoto2013} For the lattice model of Eq.\ (\ref{eq:Hlattice}), we always have $W_\pm(0) = 0$ and 
\begin{align}
  W_\pm(\pi) = \begin{cases}
      0 & \text{if } |M| > 2, \\
  \mp 1 & \text{if } -2 < M < 0, \\
  \pm 1 & \text{if } 0 < M < 2. \end{cases}
\end{align}
The strong bulk invariants are $Q_{\rm bulk} = W_-(\pi)$ so that
\begin{align}
  Q_{\rm bulk} = \begin{cases} 
     0 & \text{if } |M| > 2, \\
    +1 & \text{if } -2 < M < 0, \\
    -1 & \text{if } 0 < M < 2.
  \end{cases}
\end{align}

\begin{figure}[t]
    \centering
    \includegraphics[width=\columnwidth]{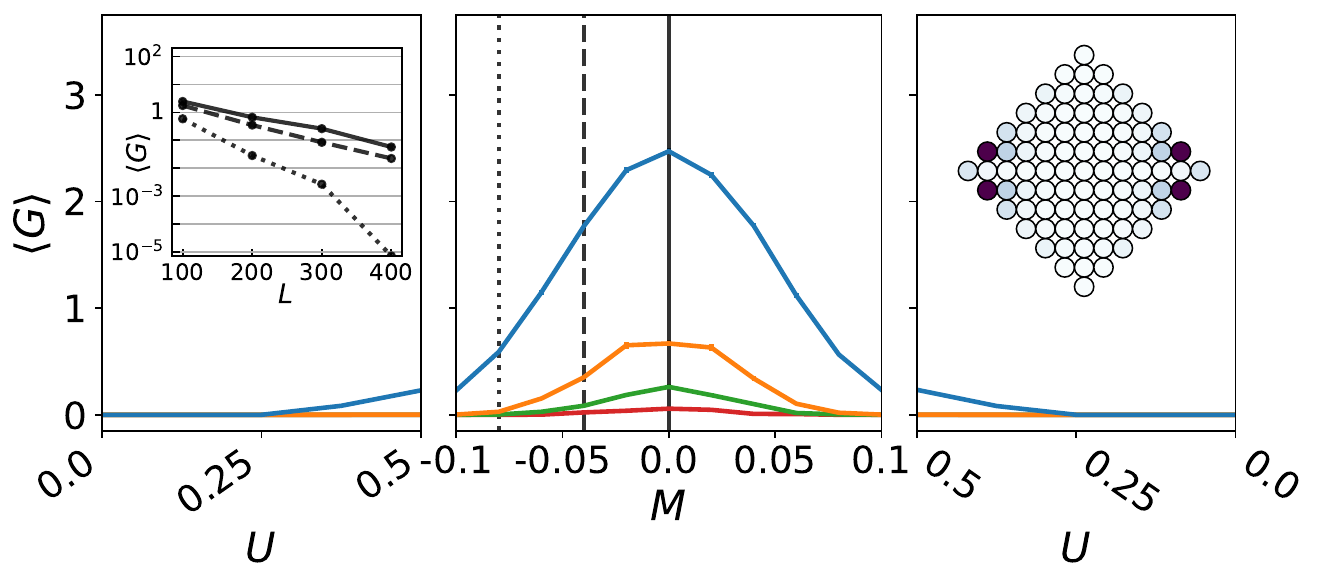}
    \includegraphics[width=\columnwidth]{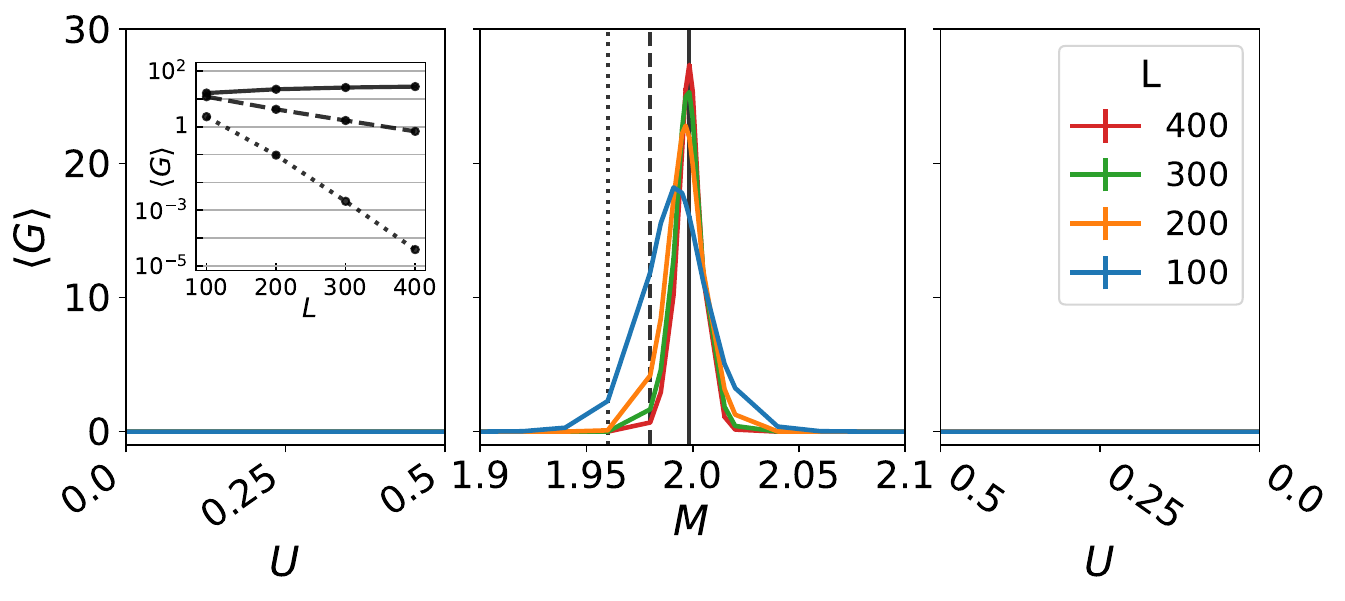}
	\caption{Disorder-averaged dimensionless two-terminal conductance $\langle G \rangle$ of an $L \times L$ disordered segment of the lattice model (\ref{eq:Hlattice}) with periodic boundary conditions in the transverse direction, with model parameters $M$ and $U$ interpolating between values that correspond to different topological phases in the absence of disorder, as described in the main text. The top panels describe an interpolation between the phases $-2 < M < 0$ and $0 < M < 2$, which are distinct in the clean limit, but not in the presence of disorder. The bottom panels describe an interpolation between phases $0 < M < 2$ and $M > 2$, which are distinct with or without disorder. Parameter values are: $M = -0.1$ ($M = 1.9$) for the top (bottom) right panels, disorder strength $U=0.5$ for the center panels, and $M = 0.1$ ($M=2.1$) for the top (bottom) left panels. The transverse coupling $\delta = 0.2$ throughout. The left insets show $\langle G \rangle$  vs.\ system size $L$ for $U = 0.5$ and for values of $M$ indicated by the vertical black lines in the center panel. The exponential decrease of $\langle G \rangle$ with $L$ for $M=0$ indicates the presence of a mobility gap at $M=0$, consistent with the absence of a topological phase transition between the phases $-2 < M < 0$ and $0 < M < 2$ in the presence of disorder. The increase of $\langle G \rangle$ with $L$ for $M=2$ signals a mobility gap closing at $M=2$, indicating a topological phase transition at that value of $M$. The top right inset illustrates the corner-state wavefunction in a square geometry with mirror-symmetric terminations for $M=0.5$ and $\delta = 0.2$.}
    \label{fig:bulk_transition}
\end{figure}

In a square geometry obtained by terminating the crystal across the diagonal lines $|x+y| = L$ and $|x-y| = L$, see Fig.\ \ref{fig:bulk_transition}, there are two mirror symmetric corners, at which there are zero-energy corner states. The corner state at the right (left) corner can be characterized with chirality $\sigma = -1$ ($\sigma = +1$) at the corner on the right (left) for both $-2<M<0$ and $0<M<2$. Both these corners have the same mirror parity $\tau = -1$ ($\tau = -1$) when for $-2<M<0$ ($0<M<2$). We use the right corner to define the boundary invariant and set $Q_{\rm boundary} = +1$ $(-1)$ for $-2<M<0$ ($0<M<2$), so that $Q_{\rm boundary} = Q_{\rm bulk}$, as required by bulk-boundary correspondence. 

{\em Disordered bulk-boundary correspondence.---} Disorder renders this model's two topological phases, which have $Q_\text{boundary} =+1$ or $Q_\text{boundary} =-1$ equivalent: The disordered invariant $\dis Q_\text{boundary} = 1$ in both cases. Hence, in the presence of disorder there is a single topological phase for $-2<M<2$ and a trivial one for $|M|>2$. 

To demonstrate the topological equivalence of the previously distinct bulk phases we numerically calculate the conductance (Fig.\ \ref{fig:bulk_transition}) of a disordered segment with periodic boundary conditions in the transverse direction. We add uncorrelated on-site (in position space) potentials 
\begin{equation}
  v_{\pm,\vr \vr'} = \sum_{\alpha=1}^{2} \sum_{\beta = 2}^{3} v_{\pm,\vr}^{\alpha, \beta} \sigma_{\alpha} \tau_{\beta}
  \delta_{\vr,\vr'}
\end{equation}
to the diagonal blocks of $H$, consisting of terms that anticommute with both $\mathcal{C}$ and $\mathcal{M}$, with weights $v_{\pm,\vr}^{\alpha,\beta}$ sampled from identical independent normal distributions with zero mean and with variance
\begin{equation}
  \langle (v_{\pm,\vr}^{\alpha, \beta})^2 \rangle = U^2.
\end{equation}
Anticommutation with $\mathcal{C}$ and $\mathcal{M}$ ensures that the disorder respects chiral symmetry, but breaks mirror symmetry (while preserving it on average). 

We compute the conductance using the software package kwant \cite{groth2014} as we interpolate between the previously distinct phases $-2<M<0$ and $0<M<2$. The interpolation is split into three parts: (i) starting from a clean system with $-2 < M < 0$, we increases disorder strength $U$ while keeping $M$ fixed, (ii) we increase $M$ to a value with $0 < M < 2$, while keeping the disorder strength $U$ fixed, and (iii) we decreases $U$ back to zero at a fixed $M$ in the range $0<M<2$. 

The conductance shows a peak at $M = 0$, where the phase boundary of the clean system was located. In the presence of disorder, this peak is, however, suppressed with increasing system size, indicating the presence of a bulk mobility gap. Disorder thus enables interpolation between $-2 < M < 0$ and $0 < M < 2$ without closing the mobility gap.

These findings are in contrast to the behavior of the conductance when we increase $M$ from a value with $0 < M < 2$ to a value $M > 2$, where we find that the conductance peak at the topological phase transition at $M=2$ increases with system size even in the presence of disorder, which is indicative of a mobility gap closing. This is consistent with the prediction that the model Eq.\ (\ref{eq:Hlattice}) is in topologically distinct phases for $M>2$ and $0< M<2$, even in the presence of disorder.

{\em Disorder-induced extrinsic transition.---} The intrinsic invariant $\dis Q_\text{boundary}$ is robust to disorder that respects the crystalline symmetry on average, as long as it does not close the bulk mobility gap. Disorder can, however, close and reopen a boundary gap and in so doing behaves as a topologically nontrivial boundary decoration. The precise configuration of anomalous boundary states may thus change upon increasing the disorder strength, but the {\em equivalence class} corresponding to the intrinsic classification does not. In other words, disorder that closes the mobility gap at the boundary, but not in the bulk, may change the extrinsic invariant $\dis N_\text{boundary}$, but it must leave the intrinsic invariant $\dis Q_\text{boundary}$ unchanged. Such a disorder-induced extrinsic phase transition is what we seek to illustrate for a lattice model in symmetry class $\text{AIII}^\mathcal{M_+}$next.

We start by considering a crystal edge normal to the reflection axis ($y$). Such an edge maps onto itself under $\mathcal{M}$. In a (minimal) topological phase it has a gapless mode, which can be described by an effective two-band edge Hamiltonian of the form
\begin{equation}
  H_{\rm edge} = -i v k_y \rho_1,
\end{equation}
where $y$ is the coordinate along the edge, and $\rho_i$ are the Pauli matrices. Chiral and mirror symmetries are represented as
\begin{align}
  H_{\rm edge}(k_y) =&\, - \rho_3 H_{\rm edge}(k_y) \rho_3 \nonumber \\ =&\,
  \rho_3 H_{\rm edge}(-k_y) \rho_3.
\end{align}
For a generic lattice termination, a differently oriented edge (with coordinate $s$ along the edge) is described by an effective edge Hamiltonian with an additional term, which gaps out the edge spectrum,
\begin{equation}
  H_{\rm edge}(k_s) = -i v k_s \rho_1 + m \rho_2.
  \label{eq:Hedge}
\end{equation}
The ``gap parameter'' $m$ in the edge theory takes opposite values for edges that are mirror images of each other. When two such edges meet at a mirror-symmetric corner, the gap parameter $m$ changes sign, leading to a zero-energy bound state at the corner.\cite{peng2017, song2017, schindler2018, langbehn2017}

Upon introducing disorder, $m$ becomes a random quantity, which depends on the coordinate $s$ and which may be characterized by the mean $\langle m \rangle \equiv \langle m(s) \rangle$ and the covariance $\langle m(s) m(s') \rangle - \langle m \rangle^2$. Importantly, disorder not only causes fluctuations of the gap function, but it may also lead to a change of the mean value $\langle m \rangle$ away from its value $m$ in the absence of disorder. (Note that mirror symmetry is broken at a generically oriented edge, so the statistical mirror symmetry of the disorder does not rule out an average effect in the edge Hamiltonian Eq.\ (\ref{eq:Hedge}).)

If the disorder is sufficiently strong, it may cause the average gap parameter $\langle m \rangle$ to undergo a sign change. A sign change of $\langle m \rangle$ corresponds to a closing and reopening of the edge mobility gap. If the disorder respects the mirror symmetry on average, the sign change of $\langle m \rangle$ (as a function of disorder strength) takes place on both sides of a mirror-symmetric corner simultaneously. Hence, there will still be a sign change of $\langle m \rangle$ (as a function of position) on both sides of the corner, so that a zero-energy corner state reappears after the disorder-induced gap closing and reopening. The two zero-energy states corresponding to the two types of domain wall at a mirror-symmetric corner have the same {\em intrinsic} invariant $\dis Q_\text{boundary}$.

To show how this scenario plays out in a lattice model, we consider a four-band lattice model, described by 
\begin{align} 
    H = &\, (M - \cos k_x - \cos k_y )\ \sigma_2\tau_0 - \dis{m} \sigma_2 \tau_1 \nonumber \\ 
    &\, \mbox{} + \sin k_x\ \sigma_1 \tau_1 + \sin k_y\ \sigma_1 \tau_3,
    \label{eq:Hlattice2}
\end{align}
where $M$ is a parameter that sets the topological phase; the term $\dis{m} \sigma_2 \tau_3$ has no effect on the bulk topology of the band structure (we assume $|\dis{m}|<|M|$), but contributes to the gap parameter $m$ in the effective edge Hamiltonian. Chiral and mirror symmetries are represented as $H(k_x,k_y) = - \sigma_3 H(k_x,k_y) \sigma_3$ and $H(k_x,k_y) = \tau_1 H(k_x,-k_y) \tau_1$, respectively. The bulk is topological for $-2<M<0$ with $W_\pm(0)=0$ and $W_\pm(\pi)=\pm1$. 

The crystal edge of the model (\ref{eq:Hlattice2}) can be described by an effective low-energy surface theory of the form (\ref{eq:Hedge}). An edge that is at an angle $\phi$ to the $x$-axis (with the bulk to the right of the edge) is characterized by edge gap parameter $m = \dis{m}\cos\phi$ and momentum $\vk_s = k_s(\cos\phi, \sin\phi)$. Mirror-related edges have angles $\phi$ and $\pi-\phi$. The Pauli matrices of the effective edge theory are related to the those of the bulk Hamiltonian (\ref{eq:Hlattice2}) as
\begin{align}
    \begin{split}
        \rho_1 &= P(\phi)\sigma_2\tau_2  P(\phi), \\
        \rho_2 &=-P(\phi)\sigma_1\tau_2  P(\phi), \\
        \rho_3 &= P(\phi)\sigma_0\tau_1' P(\phi).
    \end{split}
\end{align}
where $P(\phi) = \tfrac{1}{2}(\sigma_3\tau_1' - \sigma_0\tau_0)$ is the projector onto the two-dimensional subspace of states exponentially localized to the edge and $\tau_1' = e^{-i(\phi-\pi/2)\tau_2/2} \tau_1 e^{i(\phi-\pi/2)\tau_2/2}$. 

\begin{figure}
    \centering
    \includegraphics[width=0.8\columnwidth]{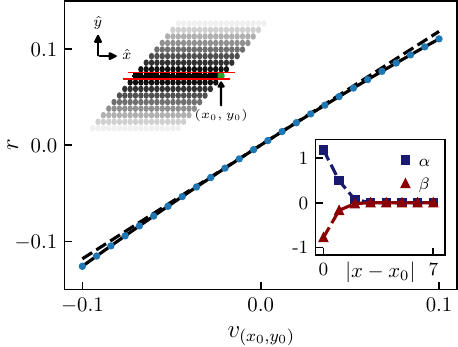}
    \caption{Main panel: Reflection amplitude $r$ vs.\ impurity potential $v_{(x_0,y_0)}$ for a position $(x_0,y_0)$ at the sample edge. The data points represent the result of a numerical calculation for the model (\ref{eq:Hlattice2}) with $M = -1.1$ and $\dis{m} = 0$. The solid curve is a quadratic fit; the dashed line is a linear reference. Top left inset: Illustration of lattice geometry used for scattering problem. A disorder potential only exists for lattice positions $(x,y_0)$ in a thin slice (between the red lines). The lattice site $(x_0,y_0)$ at the sample edge is shown in green. Bottom right inset: Coefficients $\alpha(x)$ and $\beta(x,x)$ describing the quadratic fit of the reflection amplitude $r$ to the strength $v_{(x,y_0)}$ of the impurity potential at position $(x,y_0)$ as a function of the distance $|x-x_0|$ from the crystal edge.}
    \label{fig:dis_reflection}
\end{figure}

To demonstrate that for the lattice model (\ref{eq:Hlattice}) a disorder term may indeed effectively lead to a change of the edge gap parameter, we add an on-site disorder potential (in position representation)
\begin{equation}
  V_{\vr,\vr'} = \frac{1}{2} v_{\vr} (\sigma_1 \tau_2 - \sigma_2 \tau_0) \delta_{\vr,\vr'},
\end{equation}
where the coefficients $v_{\vr}$ have identical and independent distributions with zero mean and with variance
\begin{equation}
    \langle v_{\vr}^2 \rangle = U^2.
\end{equation}
(The condition $\langle v_{\vr} \rangle = 0$ ensures that the disorder potential respects the mirror symmetry on average.)
We consider an edge with $\phi = \pi/4$ in a strip geometry, see Fig.\ \ref{fig:dis_reflection}, terminate the square lattice by truncating all bonds at the edge,  and choose $\dis{m}=0$, so that the gap parameter $m$ of the effective edge Hamiltonian vanishes in the absence of disorder. The width of the strip is chosen sufficiently large that the gapless states at the two sample edges are well separated and each edge carries a well-defined pair of counterpropagating modes at $\varepsilon = 0$ in the absence of disorder.
We then consider a scattering problem, for which the disorder is nonzero only in a slice consisting of a single row of lattice sites at coordinate $(x,y_0)$, as indicated in Fig.\ \ref{fig:dis_reflection}.
The reflection amplitude $r$ of a gapless edge mode at $\varepsilon = 0$ may be Taylor-expanded in the on-site potentials $v_{x} \equiv v_{(x,y_0)}$,
\begin{equation}
    r \simeq \sum_{x} \alpha(x) v_{x} + \sum_{x,x'} \beta(x,x') v_{x} v_{x'} + \ldots
  \label{eq:rUfit}
\end{equation}
where the summation is over all coordinates $x$ in the disordered slice, see Fig.\ \ref{fig:dis_reflection}. The functions $\alpha(x)$ and $\beta(x,x')$ can be calculated efficiently numerically using the kwant package.\cite{groth2014} Results for a strip geometry with a width of 100 sites are shown in Fig.\ \ref{fig:dis_reflection}. 

Upon disorder averaging we find that, up to quadratic order in the disorder strength $U$, the ensemble average of the reflection amplitude and its variance are
\begin{equation}
  \langle r \rangle = U^2 \sum_{\vr} \beta(x,x), \ \
  \mbox{var}\, r = U^2 \sum_{x} \alpha(x)^2.
  \label{eq:rU}
\end{equation}
To compare with the effective edge theory, we solve the analogous scattering problem for a one-dimensional Hamiltonian of the form (\ref{eq:Hedge}), with a position-dependent random gap parameter $m(s)$ with $\langle m(s) \rangle = \langle m \rangle$ and $\langle m(s)m(s')\rangle = \lambda^2\delta(s-s')$. For a disordered slice of width $\Delta L$, one then finds
\begin{equation}
  \langle r \rangle = \langle m \rangle \Delta L , \ \ \mbox{var}\, r = \lambda^2 \Delta L.
\end{equation}
Comparing with Eq.\ (\ref{eq:rU}), restoring the gap parameter $\dis{m}$, and taking into account that the effective width of the disordered slice in the lattice model (measured along the edge) is $\sqrt{2}$, we find that we can identify (for an edge with $\phi = \pi/4$ and to first order in $\dis{m}$)
\begin{align}
  \langle m \rangle =&\, \frac{1}{\sqrt{2}} \left(\dis{m} + U^2 \sum_{x} \beta(x,x) \right), \nonumber \\
  \lambda^2 =&\, \frac{1}{\sqrt{2}} U^2 \sum_{x} \alpha(x)^2,
\end{align}
confirming that the disorder term indeed leads to a change of the average edge gap parameter and, hence, can drive the edge theory through a gapless point.

\section{Conclusion}
\label{sec:conc}

Like clean topological crystalline phases, disordered TCPs, in which the crystalline symmetry is only present on average, have a rich boundary phenomenology: Higher-order disorder-robust topological states at surfaces, hinges, or corners, and obstructed atomic limits with filling anomalies. Disordered TCPs may also be in a statistical higher-order topological phase, characterized by a critical, delocalized zero-energy hinge state. Moreover, disordered topological crystalline phases obey a complete bulk-boundary correspondence: A TCP with nontrivial bulk topology either has unique higher-order boundary states or a filling anomaly. This is in contrast to the partial bulk-boundary correspondence of clean TCPs, for which knowledge of the boundary determines the bulk topology up to obstructed atomic limits without a filling anomaly. In this article, we have presented a comprehensive classification of disordered TCPs in the presence of statistical mirror, rotation and inversion symmetries. Our results provide a framework for interpreting the wide range of previous work investigating the robustness of crystalline phases to disorder.\cite{corbae_amorphous_2023, wang_structural-disorder-induced_2021, agarwala_higher-order_2020, peng_density-driven_2022, loio_third-order_2024, li_topological_2020, yang_higher-ordpeer_2021, su_disorder_2019, wang_disorder-induced_2020, hu_disorder_2021,coutant_robustness_2020,zhang_experimental_2021,franca_phase-tunable_2019,shen_disorder-induced_2024,song_delocalization_2021,peng_higher-order_2021, varjas_topological_2019, velury_topological_2021, spring_amorphous_2021, spring_isotropic_2023,marsal_obstructed_2023, wang_anderson_2024}

To arrive at a classification of disordered TCPs, we first classified intrinsic anomalous boundary signatures, {\em i.e.}, anomalous boundary states and filling anomalies that can not be obtained from perturbations localized to the crystal boundary. We then show explicitly, via the stacking construction for higher-order TCPs\cite{isobe2015,fulga2016,huang2017,khalaf2018,trifunovic2019} and via the symmetric Wannier state representation for obstructed atomic-limit TCPs,\cite{vanmiert2018} that TCPs satisfy a bulk-boundary correspondence: TCPs with the same intrinsic boundary signatures can be continuously deformed into each other without closing the bulk gap, whereas TCPs with different signatures can not. Once the bulk-boundary correspondence was established, boundary signatures could be used as bona-fide indicators of bulk topology.

Our boundary-based classification contrasts with approaches taken in the literature, where a bulk invariant of a disordered TCP is constructed by restoring translation invariance in some form and then classifying the resulting effective Hamiltonian or self-energy. To this end, Ref.\ \onlinecite{li_topological_2020} employs the self-consistent Born approximation, whereas Refs.\ \onlinecite{varjas_topological_2019,spring_amorphous_2021,spring_isotropic_2023, marsal_obstructed_2023} project the single-particle Green function of the disordered TCP onto a plane-wave basis. Such approaches may miss simplifications to the classification brought about by disorder. For example, by projection on the plane-wave basis, Ref.\ \onlinecite{spring_isotropic_2023} constructs a $\mathbb{Z}$ bulk invariant for amorphous systems with average inversion symmetry and continuous rotation in 3d, whereas the boundary in Ref.\ \onlinecite{spring_isotropic_2023} only has a $\ZZ_2$ classification. We find the same boundary classification for second-order TCPs in class $\text{A}^\mathcal{I}$, but also find that the bulks of TCPs with the same boundary states can always be continuously deformed into each other, ruling out a more refined bulk classification than the $\ZZ_2$ boundary classification.

A subtlety to topological classification in the presence of disorder is whether preserving the spectral or mobility gap is the relevant condition for topological equivalence. If two Hamiltonians can be continuously deformed into each other while keeping the spectral gap open, the mobility gap also remains open, but the opposite need not be true. That the two choices give different classifications for disordered TCPs can be seen in the example of an obstructed atomic limit: If topological equivalence is defined via the (more generous) condition that the mobility gap remain open, all atomic limits are automatically trivialized, including those with a filling anomaly. On the other hand, if topological equivalence is defined via the (more restrictive) condition that a spectral gap must be preserved, only obstructed atomic limits without filling anomaly are trivialized. The underlying reason is that the spectral-gap condition enforces that not only the Hamiltonian, but also the ground state be continuous during the deformation process. In this case a filling anomaly poses an obstruction to trivialization. On the other hand, if only a mobility gap is enforced, the ground state may change discontinuously if localized states move through the Fermi level. The classification results presented in this article are obtained for the spectral gap condition.

The TCPs considered here are of the ``strong'' type. In addition to strong TCPs, there also exist weak topological phases, which inherit their topology from lower dimensions: A weak topological phase in dimension $d$ can be continuously deformed to a stack of strong phases in dimension $< d$. An example is the weak topological insulator in three dimensions, which can be continuously deformed to a stack of two-dimensional quantum spin-Hall insulators.\cite{fu2007} An example of a weak TCP in $d=3$ is a stack of mirror-symmetric two-dimensional TCPs in class $\mbox{AIII}^{{\cal M}_+}$ (see Sec.\ \ref{sec:Ex1}), which has a flat band of zero-energy states at a mirror-symmetric hinge. Disordered weak TCPs can also be robust to disorder and may also form statistical phases similar to the disordered weak topological insulator.\cite{ringel2012} A comprehensive classification of weak TCPs is left for future work.

From the perspective of this article, clean TCPs represent configurations fine-tuned to the absence of translational symmetry breaking. Interpreted in this way, the present work contains a complete classification of ``generic'' TCPs. Each of these classes defines a symmetry protected phase of matter, with the option of undergoing \textit{phase transitions} into other phases. We hope that the classification provided below helps in navigating the multitude of arenas for critical behavior in this material class.

\acknowledgments
We would like to thank Vatsal Dwivedi, Luka Trifunovic and Anton Akhmerov for helpful discussions. This work was supported by the Deutsche Forschungsgemeinschaft (DFG, German Research Foundation) - Project Number 277101999 - CRC TR 183 (project A03).

\bibliographystyle{apsrev4-1}
\bibliography{bibfile.bib}

\appendix

\pagebreak

\section{Bulk-boundary correspondence for higher-order topological crystalline phases}
\label{app:BB}

In Secs.\ \ref{sec:Ex1} and \ref{sec:Ex2} of the main text we have seen that the presence of disorder may lift the topological distinction between certain higher-order boundary states on symmetry-invariant corners or hinges in TCPs with mirror or twofold rotation symmetry. This blurring of topological distinctions leads to a simplification of the classification of boundary states of disordered TCPs with mirror- and twofold rotation symmetries, in comparison to their clean counterparts. We here show that the classification of disordered higher-order TCPs is subject to a bulk-boundary correspondence, so that the simplification of the boundary classification implies a corresponding simplification of the bulk classification.

The existence of a bulk-boundary correspondence for disordered second-order and third-order TCPs follows from a combination of the following two observations:  (i) Disordered TCPs with different intrinsic boundary invariants $\dis Q_{\rm boundary}$ are topologically distinct, {\em i.e.}, they cannot be continuously deformed into each other without closing the bulk mobility gap; (ii) disordered TCPs with the same intrinsic boundary invariant $\dis Q_{\rm boundary}$ can always be continuously deformed into each other without closing the mobility gap. Observation (i) follows from the fact that zero-energy boundary states with nontrivial intrinsic invariant $\dis Q_{\rm boundary}$ cannot be removed by a perturbation at the crystal boundary. The proof of observation (ii) is more technical, as it makes use of the layer representation of higher-order TCPs. The discussion below specializes to the two case studies of second-order TCPs that were discussed in the main text: two-dimensional TCPs in class $\mbox{AIII}^{{\cal M}_+}$ and three-dimensional TCPs in clsas $\mbox{D}^{{\cal M}_+}$. However, the arguments can be easily generalized to other higher-order topological phases with anomalous zero-energy corner states or hinge states.

\subsection{Class \texorpdfstring{$\mbox{AIII}^{\cal{M}_+}$}{AIII\^{}M+}, \texorpdfstring{$d=2$}{d=2}}

\begin{figure}
  \centering
  \includegraphics[width=0.99\columnwidth]{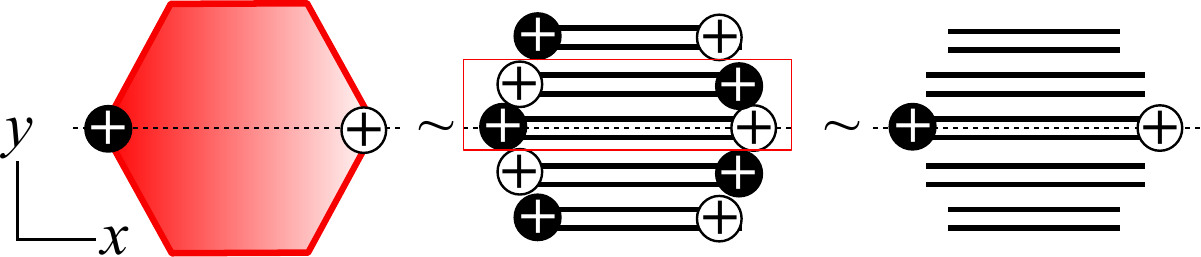}
  \caption{A second-order TCP in class $\mbox{AIII}^{{\cal M}_+}$ can be continuously deformed to a stack of one-dimensional insulating chains with alternating topological invariants, such that the corner states of the two-dimensional insulator become the end states of the one-dimensional chains. The figure shows this layer representation for a second-order insulator with a single zero-energy corner state of even mirror parity at each mirror-symmetric corner. The mirror symmetry $y \to -y$ of the stack is non-local with respect to the unit cell of the stack (red box, center panel). A weak coupling between the one-dimensional layers that respects the mirror symmetry removes the end states, except at a mirror-symmetric corner (right panel).}
    \label{fig:stacks02a}
\end{figure}

{\em Layer representation.---} A second-order TCP in class $\mbox{AIII}^{{\cal M}_+}$ can be continuously deformed to a stack of uncoupled one-dimensional band structures along $x$,\cite{isobe2015,fulga2016,huang2017,khalaf2018,trifunovic2019} shown schematically in Fig.\ \ref{fig:stacks02a}. Such a deformation may involve the addition of trivial bands, which is consistent with the rules of stable topological equivalence. Each one-dimensional layer satisfies a mirror symmetry $y \to -y$, which, seen within one layer, is a {\em local} symmetry, because mirror reflection does not affect $x$. Neighboring ``layers'' have opposite topological invariants, so that the stack as a whole has vanishing weak invariants. A perturbation that weakly couples chains at the crystal boundary, while obeying the mirror symmetry, removes all end states, with the exception of end states at a mirror-symmetric corner, which become the corner states of the two-dimensional structure. This procedure is shown schematically in Fig.\ \ref{fig:stacks02a} for a second-order TCP with a single corner state at each mirror-symmetric corner. The stacking representation is not unique, as it depends on the configuration of zero-energy states at the mirror-symmetric corners, which in turn depends on the crystal termination and may be changed by a boundary decoration.

\begin{figure}
  \centering
  \includegraphics[width=0.99\columnwidth]{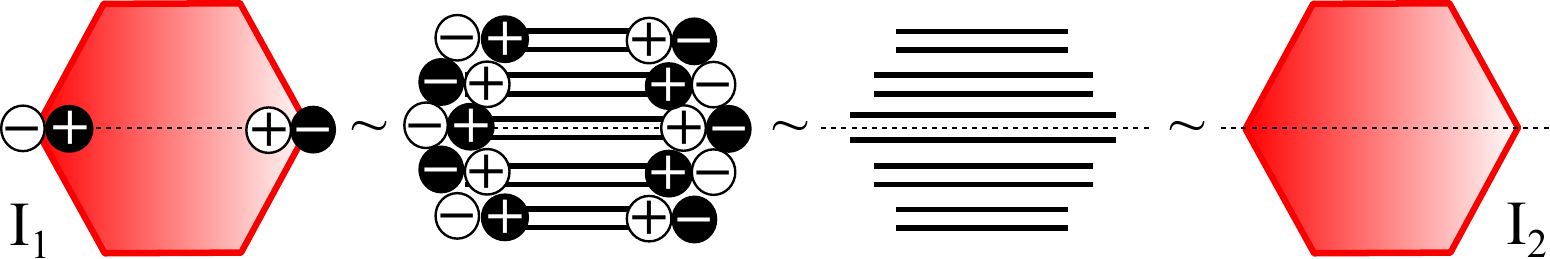}
  \caption{In the presence of disorder, class $\mbox{AIII}^{{\cal M}_+}$ insulators $\Ins_1$ and $\Ins_2$ with bulk invariants $\overline{Q}_{\rm bulk} = 2$ and $\overline{Q}_{\rm bulk} = 0$ can be continuously deformed into each other without closing the bulk mobility gap. This is achieved by choosing the boundary termination of $\Ins_1$ such that it has two zero-energy corner states of opposite chirality and opposite mirror parity, whereas $\Ins_2$ has no corner states. (Mirror parity is indicated by the sign $\pm$, positive and negative chirality is indicated by filled vs.\ open circles, respectively.) For each insulator the layer representation is used as indicated in the figure. In the presence of disorder, a continuous deformation between the one-dimensional layers that does not close the spectral gap is possible.}
    \label{fig:stacks02b}
\end{figure}

{\em Proof of bulk-boundary correspondence.---} We are now ready to prove observation (ii). We recall that for class $\mbox{AIII}^{{\cal M}_+}$ the topological invariants with and without disorder are related as $\dis{Q}_{\rm boundary} = Q_{\rm boundary} \mod 2$, see Sec.\ \ref{sec:Ex1}. Hence, it is sufficient to show that in the presence of disorder, an insulator $\Ins_1$ with $Q_{\rm boundary} = 2$ can be continuously deformed to an insulator $\Ins_2$ with $Q_{\rm boundary} = 0$. We choose the boundary terminations such that $\Ins_1$ has one zero-energy corner state with $\tau = \sigma = 1$ and one corner state with $\tau = \sigma = -1$, whereas $\Ins_2$ has no zero-energy corner states. (The labels $\sigma$ and $\tau$ refer to chirality and mirror parity, respectively.) Using the layer representation, $\Ins_1$ and $\Ins_2$ may be represented as stacks of uncoupled one-dimensional chains along $x$, see Fig.\ \ref{fig:stacks02b}. For $\Ins_1$ the layers alternate between having two zero-energy end states with $\tau = \sigma = 1$ and $\tau = \sigma = -1$ and having two zero-energy end states with $\tau = - \sigma = 1$ and $\tau = - \sigma = -1$. For $\Ins_2$ all one-dimensional chains are topologically trivial and have no zero-energy end states.

The mirror symmetry $y \to -y$ acts locally on each chain, {\em i.e.}, each layer is described by a one-dimensional Hamiltonian $h$, which satisfies the symmetry constraints
\begin{equation}
  h(k_x) = - \sigma_3 h(k_x) \sigma_3 = \tau_3 h(k_x) \tau_3.
\end{equation}
For gapped insulators with local symmetries only there is a complete bulk-boundary correspondence, because none of the symmetries is broken at the boundary. Hence, since the presence of disorder removes the topological distinction between the anomalous end states in the layer representations of $\Ins_1$ and $\Ins_2$ --- $\Ins_1$ and $\Ins_2$ have the same disorder boundary invariant $\dis{Q}_{\rm boundary}$ ---, we conclude that it must also remove the topological distinction between the bulk phases. This means that there must exist a continuous deformation from the layer representation of $\Ins_1$ to that of $\Ins_2$ that does not involve a closing of the mobility gap inside each layer.

To make the above general argument concrete, we note that in a continuum description, the one-dimensional layers may be represented by the continuum Hamiltonians
$$
  h = i v \sigma_1 \tau_0 \partial_x + \sigma_2 \tau_3 \Delta,
$$
with $\Delta > 0$ for $\Ins_1$ and $\Delta < 0$ for $\Ins_2$. A continuous deformation between $\Ins_1$ and $\Ins_2$ requires that the gap parameter $\Delta$ passes through zero. Without disorder, such a deformation involves the closing of the spectral gap. Addition of the perturbation
\begin{equation}
  h_{\rm disorder} = m [\sigma_2 \tau_1 \cos \theta(x) + \sigma_2 \tau_2 \sin \theta(x)],
\end{equation}
with $\theta(x)$ a sufficiently smooth random function of position, allows one to continuously interpolate between $\Ins_1$ and $\Ins_2$, while keeping a spectral gap of size $\gtrsim m$ open and preserving the on-site mirror symmetry on average. (If $\theta(x)$ is not a smooth function of $x$, the spectral gap may close, but the mobility gap stays open, because eigenstates of $h + h_{\rm disorder}$ are localized even if $\Delta = 0$.\cite{brouwer1998c})

\subsection{Class \texorpdfstring{D$^{{\cal M}_+}$}{D\^{}M+}, \texorpdfstring{$d=3$}{d=3} }

\begin{figure}
  \centering
  \includegraphics[width=0.99\columnwidth]{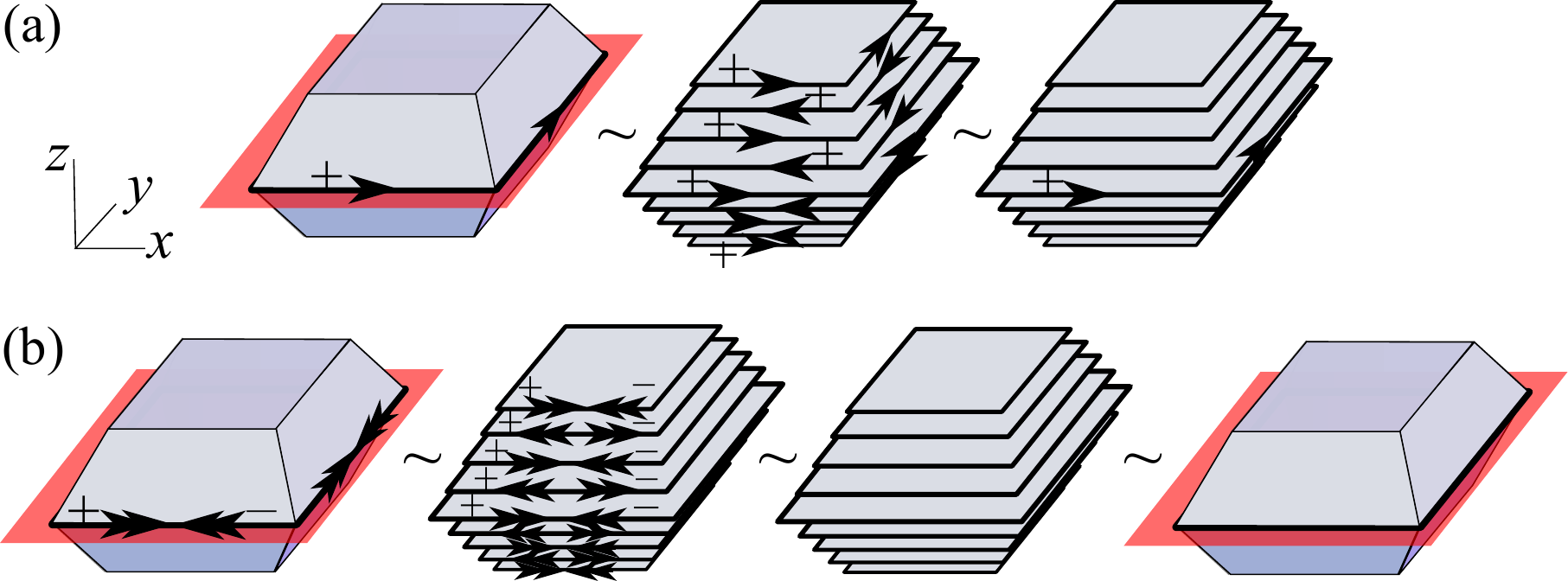}
  \caption{(a) A second-order superconductor in class $\mbox{D}^{{\cal M}_+}$ can be continuously deformed to a stack of two-dimensional topological superconductors with alternating topological invariants (left and center). The figure shows this schematically for a superconductor with a single even-parity hinge mode. A weak coupling between the layers that respects the mirror symmetry removes the chiral boundary modes, except at mirror-symmetric hinges (right). (b) Using the layer representation, one may construct a continuous deformation linking topological superconductors with $\dis{Q} = 4$ and with $\dis{Q} = 0$. Hereto, the boundary termination is chosen such that the superconductors with $\dis{Q} = 4$ has two even-parity hinge modes propagating to the right and two odd-parity hinge modes propagating to the left, whereas the superconductor with $\dis{Q} = 0$ has no hinge modes. In the presence of disorder, the two-dimensional layers with and without edge modes are topologically equivalent. \label{fig:layerD}}
\end{figure}

The above arguments can be easily carried over to other symmetry classes that allow for symmetry-invariant boundary points. As a second example, we here discuss the case of a three-dimensional topological superconductor in class $\mbox{D}^{{\cal M}_+}$. To prove the bulk-boundary correspondence, one has to show that a superconductor with $\dis{Q} = 4$ can be continuously deformed to a superconductor with $\dis{Q} = 0$. To construct such a continuous transformation we again use the layer representation. The layer representation of a second-order topological superconductors in class $\mbox{D}^{{\cal M}_+}$ consists of a stack of uncoupled two-dimensional superconductors in the $xy$ plane, whereby each layer has a (local) mirror symmetry $z \to -z$ and neighboring layers have topological invariants of opposite sign, see Fig.\ \ref{fig:layerD}(a). Figure \ref{fig:layerD}(b) shows schematically, how the topological equivalence between topological superconductors with $\dis{Q} = 4$ and with $\dis{Q} = 0$ is achieved.

\section{Bulk-boundary correspondence for obstructed atomic limits}
\label{app:dis_OAL}

Since a crystalline symmetry that leaves part of the crystal surface invariant always leads to the presence of topological states of some kind, a prerequisite for the existence of obstructed atomic-limit phases is a symmetry that acts non-locally on the entire surface, {\em i.e.} no subset of the boundary maps to itself under the symmetry. Of the order-two symmetries considered here, this condition is only satisfied by inversion ${\cal I}$. 

Atomic-limit insulators do not have anomalous boundary states, but they may have a filling anomaly, a discrepancy between the charge of an inversion-symmetric crystal and the number of unit cells in the crystal.\cite{benalcazar2019} In classes D and BDI there exists a generalized filling anomaly, which entails a discrepancy between the Pfaffian of the full inversion-symmetric crystal and the Pfaffian of the system's unit cell.\cite{chaou2025} For TCPs with inversion symmetry, a nontrivial bulk topology implies the existence of anomalous boundary states {\em or} of a (generalized) filling anomaly. Conversely, disordered TCPs without anomalous boundary states and without a filling anomaly are topologically trivial.

Here we use the real-space approach of Van Miert and Ortix\cite{vanmiert2018} to show that disorder leads to a trivialization of all atomic limit phases that do not have a filling anomaly or a generalized filling anomaly.
We elucidate our classification methodology by way of two representative examples, one for a symmetry class that is non-trivial in zero dimensions (A, AI, AII, D and BDI) and one for classes that are trivial (AIII, DIII, C, CI and CII). The adaptation of these two examples to other symmetry classes is immediate.

Before we present our two case studies, a remark on our use of the term ``atomic-limit phase'' is in order. We reserve this term for TCPs that can be obtained by continuous deformation of a trivial reference insulator, in which all electrons are in localized orbitals placed at generic Wyckoff positions, without closing the bulk mobility gap. Atomic limits constructed this way do not have anomalous boundary states, but  they may have a filling anomaly. We therefore do not consider the nontrivial phase of the Su-Schrieffer-Heeger (SSH) model (class AIII, no crystalline symmetries), an atomic limit, although it admits a basis of fully localized eigenstates. The nontrivial phase of the SSH model has anomalous zero-energy states at the two ends of the one-dimensional crystal. Continuous deformation of the nontrivial phase of the SSH model to the trivial reference phase involves closing of the mobility gap. A {\em bona fide} atomic-limit phase according to our definition is the nontrivial phase of the inversion-symmetric chain, which has a filling anomaly, but no protected end states. This example is discussed in detail in the first case study.

\subsection{Class \texorpdfstring{A$^{\mathcal{I}}$}{A\^{}I}, \texorpdfstring{$d=1$}{d=1} }
\label{app:dis_OAL_sym}

\begin{figure}
  \centering
  \includegraphics[width=0.8\columnwidth]{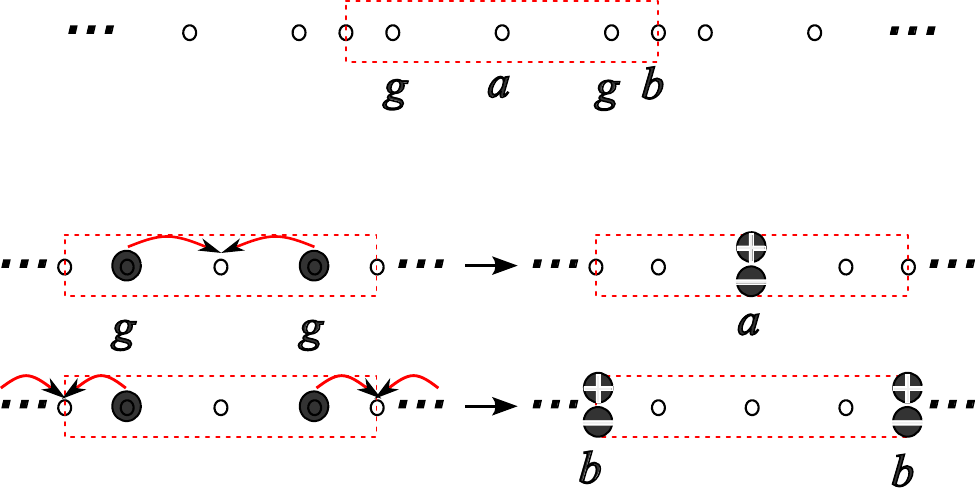}
  \caption{Top: Wyckoff positions $a$, $b$, and $g$ in an inversion symmetric one-dimensional chain. The generic Wyckoff position $g$ has multiplicity two; the special Wyckoff positions $a$ and $b$ appear once in each unit cell. Both $a$ and $b$ are inversion centers of the crystal; the two generic Wyckoff positions $g$ are mapped onto each other by inversion. Orbitals at the special Wyckoff positions have well-defined inversion parity. Bottom: A pair of states at the generic Wyckoff position $g$ may be continuously deformed to a pair of orbitals of opposite inversion parity at one of the special Wyckoff positions $a$ or $b$.}
    \label{fig:oal}
\end{figure}

A case study of an inversion-symmetric insulator in class AII was already presented in Sec.\ \ref{sec:Ex3} of the main text. We here present a different case study, class $\mbox{A}^{{\cal I}}$ in dimension $d=1$, but with a slightly more formalized construction that helps prepare us for the discussion of a case study of the class AIII in the next Subsection.

The Bloch Hamiltonian $H(k)$ of a one-dimensional inversion-symmetric TCP satisfies the symmetry constraints
\begin{equation}
  H(k) = \sigma_1 H(-k) \sigma_1 .
\end{equation}
The eigenvalue $\sigma$ of $\sigma_1$ indicates the inversion parity of the orbital. 

{\em Momentum-space picture ---} The bulk invariant is computed from the differences $D_{\sigma}(k_S) = n_{\sigma}(k_S) - n_{\rm o}$ at the two high-symmetry momenta $k_S\in \{0, \pi\}$, where $n_{\sigma}(k_S)$ is the number of occupied bands with inversion parity $\sigma$ at $k_S$ and $2 n_{\rm o}$ is the total number of occupied bands. These satisfy the constraint $D_{+}(k_S) + D_{-}(k_S) = 0$ for all $k_S$. We consider band structures without weak invariants and thus impose $D_{+}(k_S) = D_{-}(k_S) = 0$ at $k=0$. The remaining bulk topological invariant that characterizes strong phases is the integer
\begin{equation}
  Q_{\rm bulk} = D_{-}(\pi).
\end{equation}
These band structures are all of atomic-limit type. 

{\em Real-space picture ---} The atomic-limit band structures can also be obtained from a real-space picture.\cite{vanmiert2018} As a trivial reference phase, we start from an atomic-limit insulator with $n_{\rm o} + n_{\rm e}$ orbitals at each of the two generic Wyckoff positions $g$ in each unit cell. The number of occupied orbitals at each generic Wyckoff position is $n_{\rm o}$. (Since there are two generic Wyckoff positions per unit cell, the total number of electrons is $2 n_{\rm o}$ per unit cell, consistent with the $2 n_{\rm o}$ occupied bands in the momentum-space picture.) A pair of orbitals at generic Wyckoff positions can be continuously shifted to one of the special Wyckoff positions $a$ ($x=0$) or $b$ ($x=1/2$), see Fig.\ \ref{fig:oal}, where they may be rearranged a pair of orbitals at positive ($\sigma = +$) and negative ($\sigma = -$) inversion parity. 

To construct a nontrivial atomic limit, orbitals with parity $\sigma$ at one of the special Wyckoff positions $a$ or $b$ are moved from below to above the Fermi energy or vice versa. The change in the number of occupied orbitals of parity $\sigma$ at Wyckoff position $w$ is denoted $\Delta n_{\sigma}(w)$, $w \in \{a,b\}$. These changes are related to the momentum-space invariants $D_{\sigma}(k_S)$ as
\begin{align}
    D_{\sigma}(0) &= \Delta n_\sigma(a) + \Delta n_{\sigma}(b) \nonumber \\
    D_{\sigma}(\pi) &= \Delta n_\sigma(a) + \Delta n_{-\sigma}(b).
\end{align}
Over-all charge neutrality imposes the condition
\begin{equation}
  \sum_{\sigma} (\Delta n_{\sigma}(a) + \Delta n_{\sigma}(b)) = 0.
\end{equation}
In the absence of weak invariants, $D_{\sigma}(0)=0$, and the bulk invariant is given by
\begin{align}
  Q_{\rm bulk} =&\, D_{-}(\pi) \nonumber \\ =&\, \Delta n_+(b) - \Delta n_-(b).
    \label{eq:real_space_inv_1}
\end{align}

{\em Boundary invariant ---} Atomic-limit band structures with $\Delta n(b) \equiv \Delta n_{+}(b)+\Delta n_{-}(b) = 1 \mod 2$ --- so that $Q_{\rm bulk} = 1 \mod 2$ --- have a {\em filling anomaly}:\cite{benalcazar2019} If the positive lattice ions are at the unit-cell centers (Wyckoff position $a$), the parity $F$ of the total charge of an inversion-symmetric crystal equals $Q_{\rm bulk} \mod 2$. The presence of the filling anomaly leads to a fractional additional charge $(F/2) \mod 1$ at each of the two ends of the one-dimensional crystal.\cite{vanmiert2017} The $\ZZ_2$-quantity $F$ is the boundary invariant of the atomic-limit phase.

Note that the bulk-boundary correspondence is incomplete: $Q_{\rm bulk}$ determines the boundary invariant $F = Q_{\rm bulk} \mod 2$, but the boundary invariant $F$ only determines the integer $Q_{\rm bulk}$ up to a multiple of two. This underlies the fact that atomic-limit band structures cannot be distinguished via the presence of anomalous boundary charges as that not all atomic-limit band structures have fractional corner charges.

{\em Disorder ---} To obtain the bulk classification in the presence of disorder, we demonstrate that disorder trivializes all bulk phases without a boundary signature, {\em i.e.}, those with $Q_{\rm bulk} = 0 \mod 2$. This is because disorder blurs the distinction between even- and odd-parity orbitals at Wyckoff position $w=a,b$, so that the real-space invariants $\Delta n(w) = \Delta n_+(w) - \Delta n_-(w)$ are defined up to multiples of two only. It then follows from Eq.\ (\ref{eq:real_space_inv_1}) that the bulk invariant $Q_{\rm bulk}$ is defined up to a multiple of $2$, trivialising phases with $Q_{\rm bulk} = 0 \mod 2$. Furthermore, since the charge of the insulating ground state cannot change under continuous deformations, disorder that preserves the inversion symmetry on average must also preserve the filling anomaly, $\dis{F} = F$. 

We conclude that the bulk topological invariant in the presence of disorder is of $\ZZ_2$ type,
\begin{equation}
  \dis{Q}_{\rm bulk} = \dis{F} \mod 2.
\end{equation}
Disorder trivialises those phases without a boundary invariant, yielding a bulk invariant that is uniquely determined by the boundary invariants. A bulk-boundary correspondence is therefore satisfied in the presence of disorder.

{\em Spectral gap vs.\ mobility gap ---} In the above construction, the spectral gap closes upon going between different atomic limits in the absence of disorder. In the presence of disorder, a closing of the spectral gap is inevitable only if one goes between insulators with and without a filling anomaly. However, since all hybridizations between the original orbitals occur between orbitals localized at the same Wyckoff positions, the system remains an atomic-limit insulator at all times. This means that a mobility gap continues to exist, even in a transition between insulators with and without filling anomaly. 

{\em Generalisations} to other symmetry classes that are topological in zero dimensions (classes A, AI, AII, D and BDI) and to higher dimensions $d$ are straightforward. Bulk invariants for all atomic-limit phases can be formulated in terms of real-space occupation changes $\Delta n_{\sigma}(w)$ for the high-symmetry Wyckoff positions. Disorder blurs distinctions between parity sectors, rendering all atomic-limit phases equivalent, except for those that differ by the presence or absence of a (generalized) filling anomaly. For transitions from an atomic-limit phases without (generalized) filling anomaly to a phase with (generalized) filling anomaly, the average crystalline symmetry provides an obstruction to continuous deformations in the presence of disorder.

\subsection{Class \texorpdfstring{$\mbox{AIII}^{{\cal I}_-}$}{AIII\^{}I-}, \texorpdfstring{$d=1$}{d=1} }
\label{app:dis_OAL_asym}

As an example of a non-Wigner-Dyson class, we consider a one-dimensional TCP with inversion symmetry ${\cal I}$ and a unitary \textit{anti-symmetry} ${\cal C}$,
\begin{equation}
  H(k) = -\sigma_3 H(k) \sigma_3 = \sigma_1 H(-k) \sigma_1 .
\end{equation}
This symmetry class is denoted $\mbox{AIII}^{{\cal I}_-}$, where the superscript ``${\cal I}_-$'' indicates the presence of an inversion symmetry that anti-commutes with the chiral symmetry ${\cal C} = \sigma_3$. Chiral symmetry imposes that the number of occupied and unoccupied bands be equal, and since ${\cal I}$ anti-commutes with ${\cal C}$, chiral partners at energies $E$ and $-E$ must have opposite inversion-parities $\sigma$ and $-\sigma$.

{\em Momentum-space picture ---} A generic one-dimensional Hamiltonian of size $4 n_{\rm o} \times 4 n_{\rm o}$ in this symmetry class can be written as 
\begin{equation}
  H(k) = \begin{pmatrix}
  0 & A(k)^\dag \\
  A(k) & 0
  \end{pmatrix},
\end{equation}
where $A(-k)=A(k)^\dag$. First-order phases in this symmetry class are classified by the winding number 
\begin{equation}
  W = \frac{1}{2\pi} {\rm Im} \oint dk\ {\rm tr} \{ A(k)^{-1} \partial_k A(k) \}.
\end{equation}
The first-order phases are robust to disorder.

At high symmetry-points $k_S\in\{0,\pi\}$ the off-diagonal term is constrained to be Hermitian $A(k_S)=A(k_S)^\dag$. An additional bulk invariant is then given by $D(k_S) = n(k_S)-n_ {\rm o}$, where $n(k_S)$ is the number of negative eigenvalues of $A(k_S)$. Every eigenstate $\ket{v_{k_S}}$ of $A(k_S)$ can be used to build two eigenstates $( \ket{v_{k_S}}, \pm \ket{v_{k_S}})^{\rm T}$ of $H(k_S)$ with opposite inversion-parity and energy. As such $n(k_S)$ not only counts the number the number of negative eigenvalues of $A$ but also the number of positive-parity occupied states of $H(k_S)$.

The integers $D(0), D(\pi)$ and $W$ are related by a parity constraint: As ${\rm sign}\{ \det A(k_S) \} = (-1)^{n(k_S)}$, $W$ has to be odd if the $\det A(k_S)$ has opposite signs at $0$ and $\pi$. This gives
\begin{equation}
    D(0) + D(\pi) + W = 0 \mod 2.
\end{equation}
If we restrict to band structures with no weak invariants, so that $D(0)=0$, we obtain a $\ZZ^2$ bulk classification with topological invariants $W$, $D(\pi) \in\ZZ$ and the parity constraint $W = D(\pi) \mod 2$. Atomic-limit phases have $W=0$ and $D(\pi) = 0 \mod 2$.

{\em Real space picture ---} We again build a trivial reference phase by placing $n_{\rm o}$ occupied orbitals and their $n_{\rm o}$ unoccupied chiral partners at each of the two generic Wyckoff positions $g$. A pair of orbitals at the generic Wyckoff positions can be continuously moved to one of the special Wyckoff positions $a$ or $b$. There they may be rearranged into a pair of orbitals at positive ($\sigma = +$) and negative ($\sigma=-$) inversion parity, whereby there is one occupied orbital and one unoccupied orbital of each kind.

A nontrivial atomic-limit phase can be constructed from the trivial reference phase by moving states at the special Wyckoff positions through the Fermi energy. Chiral symmetry requires that for every occupied state of parity $\sigma$ moved through the Fermi energy, its unoccupied chiral partner, which has parity $-\sigma$, also be moved through the Fermi energy. We denote the change in the number of occupied orbitals of parity $\sigma$ at $w$ by $\Delta n_\sigma(w)$. Chiral symmetry enforces that $\Delta n_+(w) = -\Delta n_-(w)$. The occupation changes are related to $D(k_S)$ by
\begin{align}
    D(0)   &= \Delta n_+(a) + \Delta n_+(b) \nonumber \\
    D(\pi) &= \Delta n_+(a) + \Delta n_-(b),
\end{align}
whereas the winding number $W$ is unaffected by occupations changes of this type.
In the absence of weak invariants, $D(0)=0$. The bulk invariant is then given by
\begin{equation}
  Q_{\rm bulk} = D_{-}(\pi) = 2 \Delta n_+(a).
    \label{eq:real_space_inv_2}
\end{equation}
This always yields invariants $D(\pi) \in 2\ZZ$, consistent with what we found from the momentum-space approach.

Note that the atomic-limit phases obtained this way do not have a filling anomaly: They all have an even parity of occupied states at $w=b$. Although they are built from localized orbitals, too, TCPs with an odd number of occupied states ab $b$ cannot be obtained from the trivial reference phase by hybridization and energy shifts of localized orbitals only. Such TCPs are first-order phases with zero-energy states at the ends of a one-dimensional crystal and a nontrivial bulk invariant $W$. To obtain such phases from the trivial reference phase, orbitals located at different Wyckoff positions must be hybridized, which inevitably comes with a closing of the mobility gap.

{\em Disorder ---} Disorder trivialises all atomic-limit phases in this symmetry-class that can be obtained from the trivial reference phase by hybridization and occupation changes of localized orbitals. Because disorder locally breaks the inversion symmetry, the Hamitonian describing orbitals at the same Wyckoff position is subject to chiral symmetry only. All such Hamiltonians are topologically equivalent, leaving no room for nontrivial topology.
This conclusion is consistent with the observation that it is the existence of a filling anomaly that poses an obstruction to continuous deformation between different atomic-limit phases in the presence of disorder. For atomic-limit insulators with chiral symmetry, however, the triviality of the topological classification for $d=0$ rules out a filling anomaly.

{\em Generalisations} to other symmetry classes that are trivial in $0d$ (Classes AIII, DIII, C, CI and CII) proceed in the same way. Bulk phases can be formulated in terms of real-space invariants at special Wyckoff positions. Atomic-limits in these classes do not exhibit filling anomalies and thereby have no obstruction to trivialisation once disorder breaks the crystalline-symmetry.

\section{Second-order statistical TCP}
\label{app:statistical}

Under certain circumstances, Anderson localization can be avoided for counterpropagating zero-energy hinge states at mirror-symmetric hinges in tenfold-way classes AIII, BDI, CII, D, or DIII if the disorder respects the mirror symmetry on average. Following Ref.\ \onlinecite{fulga2014}, which identifies statistical TCPs as gapped band structures for which the boundary states are protected from localization by symmetries of the disorder ensemble, we refer to this scenario as a second-order statistical TCP. 

Instead of being exponentially localized, the zero-energy wavefunctions in a second-order statistical topological band structure have power-law correlations.
Hinge states at a finite energy $\varepsilon$ are localized, although the localization length diverges in the limit $\varepsilon \to 0$.\cite{theodorou1976,eggarter1978}
The mechanism by which zero-energy hinge states avoid localization in tenfold-way classes AIII, BDI, CII, D, or DIII applies to intrinsic as well as extrinsic anomalous hinge states.
In the classifying tables of App.\ \ref{app:classification} the possibility of a second-order statistical topological phase is indicated by an asterix.

We now discuss this scenario in more detail for a mirror-symmetric second-order topological superconductor in class D, which is the example that was presented in the main text. As in Sec.\ \ref{sec:Ex2}, we consider a hinge along $x$ that is symmetric with respect to the mirror symmetry $z \to -z$. (Because the hinge is extended in the $x$ direction, the mirror symmetry acts locally on the hinge.) We first consider a hinge with one pair of counterpropagating modes of opposite mirror parity. Without disorder, the hinge Hamiltonian $h$ satisfies the symmetry constraints
\begin{align}
  h_{\rm hinge}(k_x) =&\, -h_{\rm hinge}(-k_x)^* \nonumber \\ =&\, \tau_3 h_{\rm hinge}(k_x) \tau_3,
\end{align}
with $\tau_3$ a Pauli matrix representing mirror parity. With disorder, such a hinge may be described by a Hamiltonian of the form
\begin{equation}
  h_{\rm hinge} = i v \tau_3 \partial_x + u(x) \tau_2,
\end{equation}
where $v$ is the Fermi velocity and $u(x)$ is a random potential. The disorder is mirror symmetric, which means that random potentials $u(x)$ and $-u(x)$ occur with the same probability. Zero-energy wavefunctions can be easily constructed explicitly for this simple model,
\begin{equation}
  \psi(x) = e^{\int_{x_0}^x dx' u(x') \tau_1/v} \psi(x_0).
\end{equation}
These wavefunctions are not exponentially localized because the average $\langle u(x) \rangle = 0$.

We now turn to the general case. Hereto, we consider a hinge for which the disorder potential is set to zero for $x < x_0$, so that there are ideal left- and right-moving modes with well-defined mirror parity for $x < x_0$. We introduce matrices $p_{\rm R}$ and $p_{\rm L}$ that contain the mirror parities of the right-moving and left-moving modes on the diagonal. As discussed in Sec.\ \ref{sec:Ex2}, the topological invariants $N_{{\rm hinge},\pm}$ are differences of the numbers of left-moving and right-moving modes of the corresponding mirror parity. The presence of disorder for $x > x_0$ can be described by a reflection matrix $r(x_0,\varepsilon)$, which describes the scattering of right-moving modes at energy $\varepsilon$ into left-moving modes. Particle-hole symmetry imposes the constraint
\begin{equation}
  r(x_0,\varepsilon) = r^*(x_0,-\varepsilon).
\end{equation}
Under mirror reflection $z \to -z$, the reflection matrix changes as
\begin{equation}
  r(x_0,\varepsilon) \to p_{\rm L} r(x_0,\varepsilon) p_{\rm R}.
  \label{eq:rmirror}
\end{equation}

Exponential localization can be trivially ruled out at all energies if $N_{{\rm hinge},+} + N_{{\rm hinge},-} \neq 0$, so that it is sufficient to consider the case $N_{{\rm hinge},+} = -N_{{\rm hinge},-}$. To show that exponential localization of zero-energy eigenstates is possible only if $N_{{\rm hinge},+} = -N_{{\rm hinge},-}$ is even, we assume that the zero-energy eigenstates of $h_{\rm hinge}$ are exponentially localized and show that this leads to a contradiction if $N_{{\rm hinge},+} = -N_{{\rm hinge},-}$ is odd. If the zero-energy eigenstates of $h_{\rm hinge}$ are localized, the reflection matrix $r(x_0,0)$ is orthogonal and its determinant $\det r(x_0,0) \in \{1,-1\}$ is a topological invariant.\cite{fulga2011} Under mirror reflection $z \to -z$, $\det r(x_0,0)$ is multiplied by $\det p_{\rm R} p_{\rm L}$, see Eq.\ (\ref{eq:rmirror}). Hence, $\det r(x_0,0)$ changes sign under mirror reflection if $N_{{\rm hinge},+} = -N_{{\rm hinge},-}$ is odd. For exponentially localized states, $r(x_0,0)$ is determined by the disorder potential within a distance of a localization length $\xi$ from $x_0$; it does not depend on the details of the disorder potential for $x - x_0 \gg \xi$. Since the disorder is mirror symmetric on average, $\det r(x_0,0) = 1$ and $\det r(x_0,0) = -1$ must therefore occur with the same probability upon varying $x_0$ over distances larger than $\xi$. This, however is in contradiction with $\det r(x_0,0)$ being a topological invariant, which implies that it is independent of $x_0$. Such a contradiction does not occur if $N_{{\rm hinge},+} = -N_{{\rm hinge},-}$ is even, because $\det r(x_0,0)$ is invariant under mirror reflection in this case.

This argument can easily be generalized to show absence of exponential localization zero-energy hinge modes in class $\mbox{DIII}^{{\cal M}_{-+}}$, with the sole modification that the topological invariant of the reflection matrix $r(x_0,0)$ is its Pfaffian.
Second-order statistical insulator phases also exist for symmetry classes $\mbox{AIII}^{{\cal M}_-}$, $\mbox{BDI}^{{\cal M}_{-+}}$,  $\mbox{CII}^{{\cal M}_{+-}}$, and $\mbox{CII}^{{\cal M}_{-+}}$. In these cases, the hinge Hamiltonian has chiral symmetry. Since the mirror operation anticommutes with the chiral conjugation in all cases, the parity matrices satisfy $p_{\rm R} = - p_{\rm L}$, so that $\mbox{tr}\, r(x_0,\varepsilon) \to - \mbox{tr}\, r(x_0,\varepsilon)$ under mirror reflection. If all zero-energy states are localized, $\mbox{tr}\, r(x_0,0)$ is a topological invariant,\cite{fulga2011} which is an even integer if $N_{{\rm hinge},+} = -N_{{\rm hinge},-}$ is even and an odd integer if $N_{{\rm hinge},+} = -N_{{\rm hinge},-}$ is odd. Since $\mbox{tr}\, r(x_0,0)$ changes sign under mirror reflection, an odd-valued invariant $\mbox{tr}\, r(x_0,0)$ is incompatible with the statistical mirror symmetry of the disorder. Hence, exponential localization is possible if $N_{{\rm hinge},+} = -N_{{\rm hinge},-}$ is even, but not if $N_{{\rm hinge},+} = -N_{{\rm hinge},-}$ is odd.

\section{Classification tables}\label{app:classification}

We here present the complete boundary-resolved classification using subgroup sequences for TCPs with mirror, twofold rotation, or inversion symmetry in one, two, and three dimensions, comparing classifications without and with disorder. Boundary-resolved classifications for the non-disordered case are identical to those of Refs.\ \onlinecite{geier2018,khalaf2018b,trifunovic2019}.

The notation of the classification tables follows that of Refs.\ \onlinecite{geier2018,trifunovic2019}: The tenfold-way class is labeled with the Cartan symbol, whereas the crystalline symmetry is indicated with a superscript $\mathcal{M}$, $\mathcal{R}$, or $\mathcal{I}$ for mirror, twofold rotation, and inversion symmetry, respectively. The additional subscripts indicate, whether the crystalline symmetry commutes ($+$) or anticommutes ($-$) with the tenfold-way symmetries time-reversal $\mathcal{T}$, particle-hole $\mathcal{P}$ and/or $\mathcal{C} = \mathcal{P} \mathcal{T}$).\cite{shiozaki2014} The representations of crystalline symmetries are always chosen such that they square to one. If necessary, antiunitary crystalline symmetries, antiunitary crystalline antisymmetries, or crystalline antisymmetries can be obtained by combining a crystalline symmetry with $\mathcal{T}$, $\mathcal{P}$, or $\mathcal{C}$ respectively. In this case, a superscript $\pm$ indicates, whether $\mathcal{T}$ or $\mathcal{P}$ square to $1$ or to $-1$.  Boundary classifying groups that contain statistical topological band structures, for which zero-energy hinge states at mirror-symmetric hinges evade Anderson localization if the disorder obeys mirror symmetry on average, are indicated with an asterix. 

{\em Inversion symmetry, $d=1$, $d=2$, and $d=3$.---} Disorder does not trivialize inversion-symmetric insulators with first-order, second-order, or third-order boundary states. This is because inversion always acts non-locally on the crystal boundary, so that it has no effect on the boundary classification. Hence, the boundary-resolved classification in the presence of disorder can be easily obtained from the known classifications for the non-disordered case in Refs.\ \onlinecite{geier2018,khalaf2018b,trifunovic2019}, supplemented with classification results for disordered atomic-limit insulators. (The latter are needed, as disorder can render atomic limits that are distinct in the clean limit equivalent, thereby changing the bulk subgroup sequence.) We repeated the analysis of disordered atomic-limit as in Sec.\ \ref{sec:Ex3} for all other tenfold-way classes with an additional inversion symmetry as per the methodology laid out in App.\ \ref{app:dis_OAL}. The full classification results are given in Tables \ref{tab:inversion1d}, \ref{tab:inversion2d} and \ref{tab:inversion3d}.

{\em Mirror symmetry, $d=2$ and $d=3$.---} Reference \onlinecite{geier2018} contains extrinsic and intrinsic classifications tables for zero-energy corner states of two-dimensional mirror-symmetric insulators in the presence of a symmetry-breaking perturbation at the corner. This results in the same classification as one would obtain in the presence of a disorder potential that preserves mirror symmetry on average. These results are repeated and adapted to our notation in Tab.\ \ref{tab:mirror2d}.

In three dimensions, mirror symmetry allows for first-order phases with anomalous surface states and second-order phases with anomalous hinge modes, but no intrinsic third-order or atomic-limit phases. (There can be extrinsic zero-energy corner states at mirror-symmetric corners, however.) We repeated the analysis of Sec.\ \ref{sec:Ex2} for all other tenfold-way classes with an additional mirror symmetry as per App.\ \ref{app:statistical}. 
Subgroup group sequences for the boundary classification with and without disorder are given in Tab.\ \ref{tab:mirror3d}.

{\em Twofold rotation symmetry, $d=3$.---} In three dimensions, rotation symmetry allows for first-order, second-order, and third-order phases. First-order and second-order boundary states are robust to disorder and their classification is the same as 
for the non-disordered case.\cite{geier2018,trifunovic2019} 
For the boundary subgroup sequence for third-order phases, ${\cal D}_3$ and ${\cal D}_3''$ are identical to the boundary classification groups ${\cal D}_2$ and ${\cal D}_2'$ of second-order phases with mirror symmetry ${\cal M}$ in dimension $d=2$, respectively, see Tab.\ \ref{tab:mirror2d}. Since the effect of disorder is local, the same applies to the boundary classification groups $\dis{\mathcal{D}}_3$ and $\dis{\mathcal{D}}''_3$ in the presence of disorder, which are equal to the groups $\dis{\mathcal{D}}_2$ and $\dis{\mathcal{D}}'_2$ of the boundary classification of disordered second-order phases with mirror symmetry ${\cal M}$ in dimension $d=2$.

In all but five symmetry classes, the boundary classification group $\mathcal{D}'_3$, which classifies corner modes that can be obtained from decorations on surfaces or hinges, is equal to ${\cal D}_3''$, the group classifying corner modes that can be obtained from decorating hinges only.\cite{trifunovic2019} This relation remains valid in the presence of disorder, {\em i.e.}, for these symmetry classes one has $\dis{\mathcal{D}}'_3 = \dis{\mathcal{D}}''_3$. The five exceptions to this rule are $\text{A}^\mathcal{P^+R}$, $\text{D}^\mathcal{R_+}$, $\text{DIII}^\mathcal{R_{++}}$, $\text{DIII}^\mathcal{R_{-+}}$, and $\text{AII}^\mathcal{CR_-}$. These are symmetry classes that support ``separable'' surface decorations, which support a single corner mode. In Ref.\ \onlinecite{trifunovic2019} it was found for these symmetry classes $\mathcal{D}'_3 = \mathcal{D}_3$. Because disorder only acts locally, the same equality applies in the presence of disorder, so that one also has $\dis{\mathcal{D}}'_3 = \dis{\mathcal{D}}_3$ for these five symmetry classes in the presence of disorder.
The complete subgroup group sequences for the bulk and boundary classifications with and without disorder are given in Table \ref{tab:rotation3d}.

\vfill\eject


\begin{table}[H]
\begin{tabular*}{\linewidth}{c@{\extracolsep{\fill}} cc}
 \hline\hline
 Class &
 \multicolumn{2}{ c }{Bulk classification} \\
 &
 $\mathcal{K}' \!\subseteq\! \mathcal{K}$&
 $\dis{\mathcal{K}}' \!\subseteq\! \dis{\mathcal{K}}$ \\
 \hline \hline
 $\text{A}^\mathcal{I}$         & $ \ZZ   \subseteq \ZZ     $ & $ \ZZ_2 \subseteq \ZZ_2           $ \\
 $\text{AIII}^\mathcal{I_+}$    & $ -                       $ & $ -                               $ \\
 \hline        
 $\text{A}^\mathcal{CI}$        & $ -                       $ & $ -                               $ \\
 $\text{AIII}^\mathcal{I_-}$    & $ \ZZ   \subseteq \ZZ^2   $ & $ 0     \subseteq \ZZ             $ \\
 \hline        
 $\text{A}^\mathcal{T^+I}$      & $ \ZZ_2 \subseteq \ZZ_2   $ & $ \ZZ_2 \subseteq \ZZ_2           $ \\
 $\text{AIII}^\mathcal{T^+I_+}$ & $ \ZZ_2 \subseteq \ZZ_2   $ & $ -                               $ \\
\hline \hline
\end{tabular*}
\caption{Classification of inversion symmetric phases in one dimension. \label{tab:inversion1d}}
\end{table}


\begin{table}[H]
\begin{tabular*}{\linewidth}{c@{\extracolsep{\fill}} cc  cc }
 \hline\hline
 Class & 
 \multicolumn{2}{ c }{ Bulk classification} &
 \multicolumn{2}{ c }{ 2\textsuperscript{nd} order boundary }\\
 &
 $\mathcal{K}'' \!\subseteq\! \mathcal{K}' \!\subseteq\! \mathcal{K}$&
 $\dis{\mathcal{K}}'' \!\subseteq\! \dis{\mathcal{K}}' \!\subseteq\! \dis{\mathcal{K}}$ &
 $\mathcal{D}_2' \!\subseteq\! \mathcal{D}_2$ &
 $\dis{\mathcal{D}}_2' \!\subseteq\! \dis{\mathcal{D}}_2$\\
 \hline \hline 
 $\text{A}^\mathcal{M}$         & $-                                    $ & $ -                                 $ & $ -                      $ & $ -                   $ \\
 $\text{AIII}^\mathcal{M_+}$    & $ 0 \subseteq \ZZ   \subseteq \ZZ     $ & $ 0 \subseteq \ZZ_2 \subseteq \ZZ_2 $ & $ \ZZ    \subseteq \ZZ^2 $ & $ 2\ZZ \subseteq \ZZ  $ \\
 \hline                                                                                                                                                                                                   
 $\text{A}^\mathcal{CM}$        & $ 0 \subseteq \ZZ   \subseteq \ZZ^2   $ & $ 0 \subseteq 0     \subseteq \ZZ   $ & $ 0      \subseteq \ZZ   $ & $ -                   $ \\
 $\text{AIII}^\mathcal{M_-}$    & $-                                    $ & $ -                                 $ & $ -                      $ & $ -                   $ \\
 \hline                                         
 $\text{A}^\mathcal{T^+M}$      & $ 0 \subseteq 0     \subseteq \ZZ     $ & $ 0 \subseteq 0     \subseteq \ZZ   $ & $ -                      $ & $ -                   $ \\
 $\text{AIII}^\mathcal{T^+M_+}$ & $ 0 \subseteq \ZZ_2 \subseteq \ZZ_2   $ & $ 0 \subseteq \ZZ_2 \subseteq \ZZ_2 $ & $ 2\ZZ   \subseteq \ZZ   $ & $ 2\ZZ \subseteq \ZZ  $ \\
\hline \hline
\end{tabular*}
\caption{Classification of mirror symmetric phases in two dimensions. \label{tab:mirror2d}}
\end{table}
                                                                                              

\begin{table}[H]
\begin{tabular*}{\linewidth}{c@{\extracolsep{\fill}} cc  c }
 \hline\hline
 Class &
 \multicolumn{2}{ c }{ Bulk classification} &
 \multicolumn{1}{ c }{ 2\textsuperscript{nd} order boundary }\\
 &
 $\mathcal{K}'' \!\subseteq\! \mathcal{K}' \!\subseteq\! \mathcal{K}$&
 $\dis{\mathcal{K}}'' \!\subseteq\! \dis{\mathcal{K}}' \!\subseteq\! \dis{\mathcal{K}}$ &
 $\dis{\mathcal{D}}_2' \!\subseteq\! \dis{\mathcal{D}}_2$\\
 \hline \hline
 $\text{A}^\mathcal{I}$         & $ \ZZ   \subseteq \ZZ    \subseteq  \ZZ^2   $ & $ \ZZ_2 \subseteq \ZZ_2  \subseteq  \ZZ\times\ZZ_2   $ & $ -                     $ \\
 $\text{AIII}^\mathcal{I_+}$    & $ -                                         $ & $ -                                                  $ & $ -                     $ \\
 \hline                                                                                                                                                                                                                                           
 $\text{A}^\mathcal{CI}$        & $ -                                         $ & $ -                                                  $ & $ -                     $ \\
 $\text{AIII}^\mathcal{I_-}$    & $ 2\ZZ  \subseteq \ZZ    \subseteq  \ZZ     $ & $ 0     \subseteq \ZZ_2  \subseteq  \ZZ_2            $ & $ 2\ZZ   \subseteq \ZZ  $ \\
 \hline         
 $\text{A}^\mathcal{T^+I}$      & $ \ZZ_2 \subseteq \ZZ_2  \subseteq  \ZZ_2   $ & $ \ZZ_2 \subseteq \ZZ_2  \subseteq  \ZZ_2            $ & $ -                     $ \\
 $\text{AIII}^\mathcal{T^+I_+}$ & $ -                                         $ & $ -                                                  $ & $ -                     $ \\
\hline \hline
\end{tabular*}
\caption{Classification of inversion symmetric phases in two dimensions. \label{tab:inversion2d}}
\end{table}


\begin{table*}
\begin{tabular*}{\linewidth}{c@{\extracolsep{\fill}} cc  cc cc }
 \hline\hline
 Class & 
 \multicolumn{2}{ c }{ Bulk classification} &
 \multicolumn{2}{ c }{ 3\textsuperscript{rd} order boundary } &
 \multicolumn{2}{ c }{ 2\textsuperscript{nd} order boundary }\\
 &
 $\mathcal{K}''' \!\subseteq\!\mathcal{K}'' \!\subseteq\! \mathcal{K}' \!\subseteq\! \mathcal{K}$&
 $\dis{\mathcal{K}}''' \!\subseteq\! \dis{\mathcal{K}}'' \!\subseteq\! \dis{\mathcal{K}}' \!\subseteq\! \dis{\mathcal{K}}$ &
 $\mathcal{D}_3'' \!\subseteq\! \mathcal{D}_3' \!\subseteq\! \mathcal{D}_3$ &
 $\dis{\mathcal{D}}_3'' \!\subseteq\! \dis{\mathcal{D}}_3' \!\subseteq\! \dis{\mathcal{D}}_3$ &
 $\mathcal{D}_2' \!\subseteq\! \mathcal{D}_2$ &
 $\dis{\mathcal{D}}_2' \!\subseteq\! \dis{\mathcal{D}}_2$\\
 \hline \hline 
 $\text{A}^\mathcal{M}$         & $ 0 \subseteq 0 \subseteq \ZZ   \subseteq \ZZ   $ & $ 0 \subseteq 0     \subseteq \ZZ_2 \subseteq \ZZ_2           $ & $ -                                        $ & $ -                                  $ & $ \ZZ   \subseteq \ZZ^2 $ & $ 2\ZZ    \subseteq \ZZ  $ \\
 $\text{AIII}^\mathcal{M_+}$    & $ -                                             $ & $ -                                                           $ & $ \ZZ^2  \subseteq \ZZ^2  \subseteq \ZZ^2  $ & $ \ZZ  \subseteq \ZZ  \subseteq \ZZ  $ & $ -                     $ & $ -                      $ \\
 \hline                                                                                                                                                                                                                                                                                                
 $\text{A}^\mathcal{CM}$        & $ -                                             $ & $ -                                                           $ & $ \ZZ    \subseteq \ZZ    \subseteq \ZZ    $ & $ -                                  $ & $ -                     $ & $ -                      $ \\
\hline\hline                                                                                                                                                                                                                                                
\end{tabular*}
\caption{Classification of mirror symmetric phases in three dimensions. The subgroup relation $2 \ZZ \subseteq \ZZ \times \ZZ_2$ for the disordered boundary classifying groups of class $\mbox{D}^{{\cal M}_+}$ corresponds to the embedding $2n \to (2n,n \mod 2)$, $n \in \ZZ$. The superscript $*$ indicates a statistical higher-order phase. \label{tab:mirror3d}}
\end{table*} 


\begin{table*}
\begin{tabular*}{\linewidth}{c@{\extracolsep{\fill}} cc cc c }
 \hline\hline
 Class & 
 \multicolumn{2}{ c }{ Bulk classification} &
 \multicolumn{2}{ c }{ 3\textsuperscript{rd} order boundary } &
 \multicolumn{1}{ c }{ 2\textsuperscript{nd} order boundary }\\
 &
 $\mathcal{K}''' \!\subseteq\!\mathcal{K}'' \!\subseteq\! \mathcal{K}' \!\subseteq\! \mathcal{K}$&
 $\dis{\mathcal{K}}''' \!\subseteq\! \dis{\mathcal{K}}'' \!\subseteq\! \dis{\mathcal{K}}' \!\subseteq\! \dis{\mathcal{K}}$ &
 $\mathcal{D}_3'' \!\subseteq\! \mathcal{D}_3' \!\subseteq\! \mathcal{D}_3$ &
 $\dis{\mathcal{D}}_3'' \!\subseteq\! \dis{\mathcal{D}}_3' \!\subseteq\! \dis{\mathcal{D}}_3$ &
 $\dis{\mathcal{D}}_2' \!\subseteq\! \dis{\mathcal{D}}_2$\\
 \hline \hline 
 $\text{A}^\mathcal{R}$         & $ -                                                   $ & $ -                                                         $ & $ -                                         $ & $ -                                    $ & $ -                    $ \\
 $\text{AIII}^\mathcal{R_+}$    & $ 0 \subseteq \ZZ   \subseteq \ZZ    \subseteq \ZZ^2  $ & $ 0 \subseteq \ZZ_2 \subseteq \ZZ_2 \subseteq \ZZ\times\ZZ_2$ & $ \ZZ   \subseteq \ZZ     \subseteq \ZZ^2   $ & $ 2\ZZ \subseteq 2\ZZ \subseteq \ZZ    $ & $ -                    $ \\
 \hline                                                                                                                                                                                                                                                                                                                        
 $\text{A}^\mathcal{CR}$        & $ 0 \subseteq 2\ZZ  \subseteq \ZZ    \subseteq \ZZ    $ & $ 0 \subseteq 0     \subseteq \ZZ_2 \subseteq \ZZ_2         $ & $ 0     \subseteq 0       \subseteq 2\ZZ    $ & $ -                                    $ & $ 2\ZZ \subseteq \ZZ   $ \\
\hline\hline                  
\end{tabular*}
\caption{Classification of rotation symmetric phases in three dimensions. \label{tab:rotation3d}}
\end{table*}


\begin{table*}
\begin{tabular*}{\linewidth}{c@{\extracolsep{\fill}} cc  c c }
 \hline\hline
 Class & 
 \multicolumn{2}{ c }{ Bulk classification} &
 \multicolumn{1}{ c }{ 3\textsuperscript{rd} order boundary } &
 \multicolumn{1}{ c }{ 2\textsuperscript{nd} order boundary }\\
 &
 $\mathcal{K}''' \!\subseteq\!\mathcal{K}'' \!\subseteq\! \mathcal{K}' \!\subseteq\! \mathcal{K}$&
 $\dis{\mathcal{K}}''' \!\subseteq\! \dis{\mathcal{K}}'' \!\subseteq\! \dis{\mathcal{K}}' \!\subseteq\! \dis{\mathcal{K}}$ &
 $\dis{\mathcal{D}}_3'' \!\subseteq\! \dis{\mathcal{D}}_3' \!\subseteq\! \dis{\mathcal{D}}_3$ &
 $\dis{\mathcal{D}}_2' \!\subseteq\! \dis{\mathcal{D}}_2$\\
 \hline \hline 
 $\text{A}^\mathcal{I}$         & $ 2\ZZ \subseteq 2\ZZ  \subseteq \ZZ   \subseteq \ZZ     $ & $ \ZZ_2 \subseteq \ZZ_2 \subseteq \ZZ_4 \subseteq \ZZ_4          $ & $ -                                   $ & $ 2\ZZ  \subseteq \ZZ    $ \\
 $\text{AIII}^\mathcal{I_+}$    & $ -                                                      $ & $ -                                                              $ & $ -                                   $ & $ -                      $ \\
 \hline                                                                                                                                                                                                                                                      
 $\text{A}^\mathcal{CI}$        & $ -                                                      $ & $ -                                                              $ & $ -                                   $ & $ -                      $ \\
 $\text{AIII}^\mathcal{I_-}$    & $ 2\ZZ \subseteq \ZZ   \subseteq \ZZ   \subseteq \ZZ^2   $ & $ 0     \subseteq \ZZ_2 \subseteq \ZZ_2 \subseteq \ZZ\times\ZZ_2 $ & $ 2\ZZ \subseteq 2\ZZ  \subseteq \ZZ  $ & $ -                      $ \\
\hline \hline
\end{tabular*}
\caption{Classification of inversion symmetric phases in three dimensions. \label{tab:inversion3d}}
\end{table*}

\end{document}